\UseRawInputEncoding
\documentclass[%
 reprint,
 amsmath,amssymb,
 aps,
floatfix,
]{revtex4-2}

\usepackage{graphicx}
\usepackage{dcolumn}
\usepackage{bm}
\usepackage[usenames,dvipsnames]{color}
\usepackage[svgnames]{xcolor}
\usepackage{float}
\begin{document}

\preprint{APS/123-QED}

\title{Dynamical black hole in a bouncing universe}

\author{Daniela P\'erez}
\email{danielaperez@iar.unlp.edu.ar}

 \affiliation{%
 Instituto Argentino de Radioastronom\'ia (IAR, CONICET/CIC/UNLP), C.C.5, (1894) Villa Elisa, Buenos Aires, Argentina}%

\author{Santiago E. Perez Bergliaffa}%


\affiliation{Departamento de F\'isica Te\'orica, Instituto de F\'isica,
Universidade do Estado de Rio de Janeiro,
CEP 20550-013, Rio de Janeiro, Brazil
}%
\author{Gustavo E. Romero}
\altaffiliation[Also at ]{Facultad de Ciencias Astron\'omicas y Geof\'isicas, Universidad Nacional de La Plata, Paseo del Bosque s/n, 1900 La Plata, Buenos Aires, Argentina}
\affiliation{%
Instituto Argentino de Radioastronom\'ia (IAR, CONICET/CIC/UNLP), C.C.5, (1894) Villa Elisa, Buenos Aires, Argentina }%


\date{\today}

\begin{abstract}
We analyze the causal structure of McVittie spacetime for a classical bouncing cosmological model. In particular, we compute the trapping horizons of the metric and integrate the trajectories of radial null geodesics before, during, and after the bounce takes place. In the contracting phase up to the occurrence of the bounce, a dynamical black hole is present. When the universe reaches a certain minimum scale, the trapping horizons disappear and the black hole ceases to exist. After the bounce, the central weak singularity becomes naked. In the expanding phase, for large positive values of the cosmic time, the behaviour of null geodesics  indicates that the solution contains a black hole. These results suggest that neither a contracting nor an expanding universe can accommodate a black hole at all times.
\end{abstract}

\maketitle


\section{Introduction}\label{sec:1}

Our current understanding of the evolution of the Universe is expressed in the so-called $\Lambda$CDM model that includes gravity as described by the Einstein's field equations with a cosmological term, the standard model of particle physics, a component of cold dark matter, and a hot initial phase. This model is frequently complemented with an inflationary stage that would have occurred immediately before the grand unification epoch. Inflation requires additional physics beyond the standard model. With or without inflation the standard cosmological model is singular \cite{bor+03}. An initial cosmological singularity is a very undesirable feature that points out an irreparable deficiency in the  representation of the underlying physical processes \cite{rom13}. Bouncing cosmologies try to offer an alternative to overcome such problems. 

There is a wide variety of proposals, either classical or quantum, for a cosmological bounce \cite{Novello2008}. In all these models the universe starts from a very diluted phase and proceeds to contract. The contraction then smoothly evolves into a bounce that leads to the current phase of expansion as described by the $\Lambda$CDM model. These models solve the problem of the initial cosmological singularity. As the cosmic fluid contracts most structure is erased 
and the universe becomes smooth \cite{Rees1969}. Black holes, however, might survive the bounce and play some role in the subsequent expanding universe \cite{sik+91,car+11,Clif+2017}. Since a black hole is essentially a region of spacetime with particular curvature, the overall contraction and expansion of space during the bounce should have a global-to-local effect upon its horizons. The whole process is dynamical, and hence cannot be investigated using the standard static solutions. 

An exact solution for a central inhomogeneity in a cosmological setting was first found by McVittie long ago. It is now clear that such a solution describes a black hole. McVittie solution has been investigated, so far, for standard prescriptions of the scale factor of the universe. In this work we extend the research to models that allow for a bounce. We discuss whether the solution includes a black hole before the bounce and what happens with the horizons along the cosmological history of a bouncing universe. The results we have found, we hope, will help to obtain a better understanding of both the McVittie solution and the fate of a dynamical black hole through a cosmic bounce.   

\section{Dynamical spacetimes}

\subsection{McVittie spacetime}

In 1933, McVittie \cite{mcv33} discovered an exact solution of Einstein's field equations that describes an inhomogeneity embedded in a Friedmann-Lema\^{i}tre-Robertson–Walker (FLRW) cosmological background. In isotropic coordinates $(t,r,\theta,\phi)$ its line element takes the form:
\begin{eqnarray}\label{mcv-iso}
ds^{2} & = &  - \frac{\left(1 - \frac{m_0}{2 r a(t)}\right)^{2}}{\left(1 + \frac{m_0}{2 r a(t)}\right)^{2}}  dt^{2}
 +  a^{2}(t) \left(1 + \frac{m_0}{2 r a(t)}\right)^{4} \nonumber \\
 & \times & \left[dr^{2} + r^{2} \left( d\theta^{2} +  \sin^{2}{\theta} d\phi^{2}\right)\right].
\end{eqnarray}
 Here, $a(t)$ is the scale factor of the background cosmological model and $m_0$ is a non-negative constant. Setting $a(t) \equiv 1$,  Eq. \eqref{mcv-iso} reduces to the Schwarzschild line element in isotropic coordinates, and in the limit $m_0 \rightarrow 0$ the FLRW metric is recovered. For the upcoming discussion, it is convenient to express Eq. \eqref{mcv-iso} in terms of $R$, the areal radius coordinate \footnote{The areal radius coordinate $R$ is defined by $R:= \sqrt{\mathcal{A}/4\pi}$, where $\mathcal{A}$ is the area of the 2-sphere of symmetry, and where
\begin{equation}
 d^{2}\Omega_{(2)} := {d\theta}^{2} + {\sin{\theta}}^{2}  {d\phi}^{2},
\end{equation}
is the line element on the unit 2-sphere \cite{far15}.}:
\begin{equation}\label{areal-radius}
R \equiv a(t) r   \left(1 + \frac{m_0}{2 r a(t)}\right)^{2}.
\end{equation}
Using this expression, the McVittie line element can be written as
\begin{eqnarray}\label{mcv}
 ds^{2}  & = & - f(t,R) dt^{2} - \frac{2 H(t) R}{\sqrt{1- 2m_0/R}} dt dR + \frac{dR^{2}}{1- 2m_0/R}\nonumber\\
& +&  R^{2}\left( d\theta^{2} +  \sin^{2}{\theta} d\phi^{2}\right),
\end{eqnarray}
where
\begin{equation}
f(t,R) \equiv 1- 2m_0/R - H(t)^2 R^2.   
\end{equation}
Here, $H(t) \equiv \dot{a}(t)/a(t)$ is the Hubble factor corresponding to the
background cosmological model.  There are 
two key assumptions 
on which 
McVittie's solution is based. The first one is that the matter represented in the field equations is
described by a perfect fluid with density $\rho$ and isotropic pressure $p$. The second one is that the fluid is at rest with respect to the chosen reference frame. There are no additional hypothesis regarding the properties of the matter that sources the geometry. In particular, an equation of state, for instance of the form $p = p(\rho)$, is not assumed. The relation between $\rho$ and $p$ is obtained \textit{a posteriori} by solving Einstein field equations when the scale function $a(t)$ is specified. For further details on the derivation of the McVittie solution see Refs. \cite{car+09,nol98,nol99a}.

There has been a long debate in the literature about the physical interpretation of the McVittie spacetime, focused mainly 
on deciding
whether the solution characterizes a black hole in an expanding universe or not.  
A recent series of works \cite{nol98,nol99a,nol99b,kal+10,lak+11}  has been crucial to 
establish that 
the McVittie metric represents a dynamical black hole embedded in a cosmological background. The details of the solution and its possible analytical extension depend of
the behaviour of
$H(t)$ for $t\rightarrow\infty$ \cite{lak+11}. The solution  
displays a curvature singularity at $R=2m_0$ for finite values of $t$, as evidenced by the Ricci scalar, given by: 
\begin{equation}
{\cal R}=12H^2+\frac{6\dot H}{\sqrt{1-\frac{2m_0}{R}}}.
\end{equation}
The singularity is spacelike and, as shown by Nolan \cite{nol99b}, gravitationally weak \cite{tip77}.

The key feature associated with black holes is the presence of an event horizon, i.e. a boundary between two regions of spacetime that are causally disconnected. Events inside the black hole are separated from events in the global external future of spacetime. In dynamical spacetimes, in order to identify the event horizon, we would need to know the entire spacetime manifold to future infinity, which is impossible if the metric is not completely known.


In order to determine whether a black hole is embedded in a dynamical background, a full analysis of the causal structure of the spacetime is necessary. This includes studying the existence of trapping horizons, the determination of regular trapped and anti-trapped regions, and the computation of the trajectories of ingoing and outgoing radial null geodesics. In what follows, we briefly review the main results of Refs. \cite{nol98,nol99a,nol99b,kal+10,lak+11} that prove the 
assertion that the McVittie spacetime describes a black hole in a cosmological environment. 

The trapping horizons of a spacetime are defined as the surfaces where null geodesics change their focusing properties \cite{hay94}. Mathematically, this kind of horizon is determined by the condition
\begin{equation}
 \theta_{\mathrm{in}} \theta_{\mathrm{out}} = 0, 
\end{equation}
where $\theta_{\mathrm{in}}$ stands for the expansion of ingoing radial null geodesics while $\theta_{\mathrm{out}}$ denotes the expansion of outgoing radial null geodesics, respectively. Regions where $\theta_{\mathrm{in}} \theta_{\mathrm{out}} < 0$ are called \textit{regular}. In the opposite case, $\theta_{\mathrm{in}} \theta_{\mathrm{out}} > 0$, the region is called \textit{anti-trapped} if $\theta_{\mathrm{in}} > 0$ and $\theta_{\mathrm{out}} >0$, and \textit{trapped} if $\theta_{\mathrm{in}} < 0$ and $\theta_{\mathrm{out}} < 0$. In this section, we are only considering expanding spacetimes, i.e. $H(t) > 0$. The trapping horizons for the McVittie metric occur for  $\theta_{\mathrm{in}} = 0$, and null outgoing geodesics are always expanding ($\theta_{\mathrm{out}} > 0$) \cite{kal+10,lak+11}.

\begin{figure}[t]
\includegraphics[width=8cm]{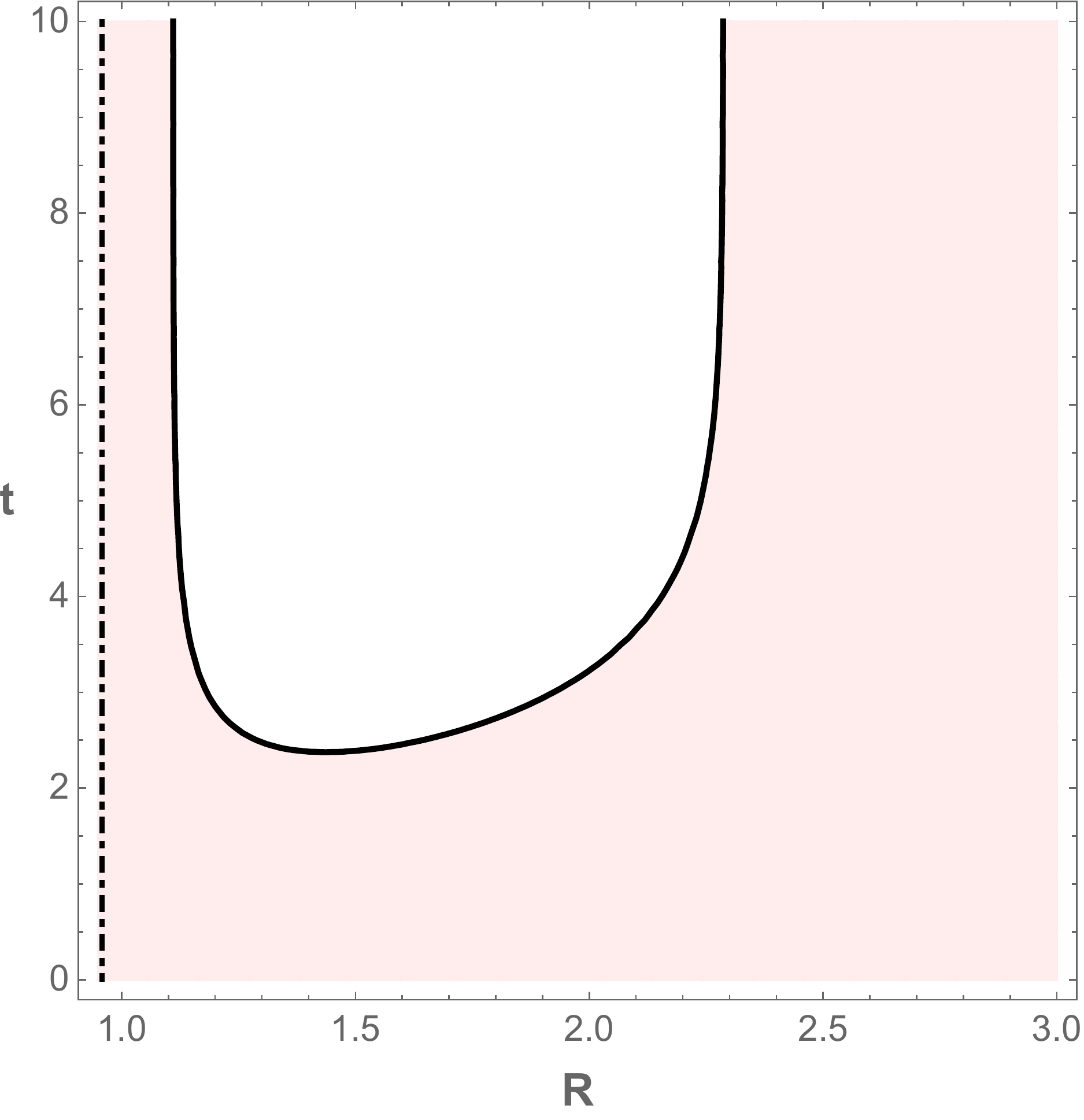}
\caption{\label{fig:1} The black line indicates the location of the trapping horizons in McVittie spacetime for the $\Lambda$CDM model. The white zone corresponds to the regular region while the light pink corresponds to the anti-trapped region, respectively. The dot dashed line denotes the location of the singularity. Here, we fixed $m_0 = 0.479$ and $H_0= 1/3$ as an example.}
\end{figure}

\begin{figure}[t]
\includegraphics[width=8cm]{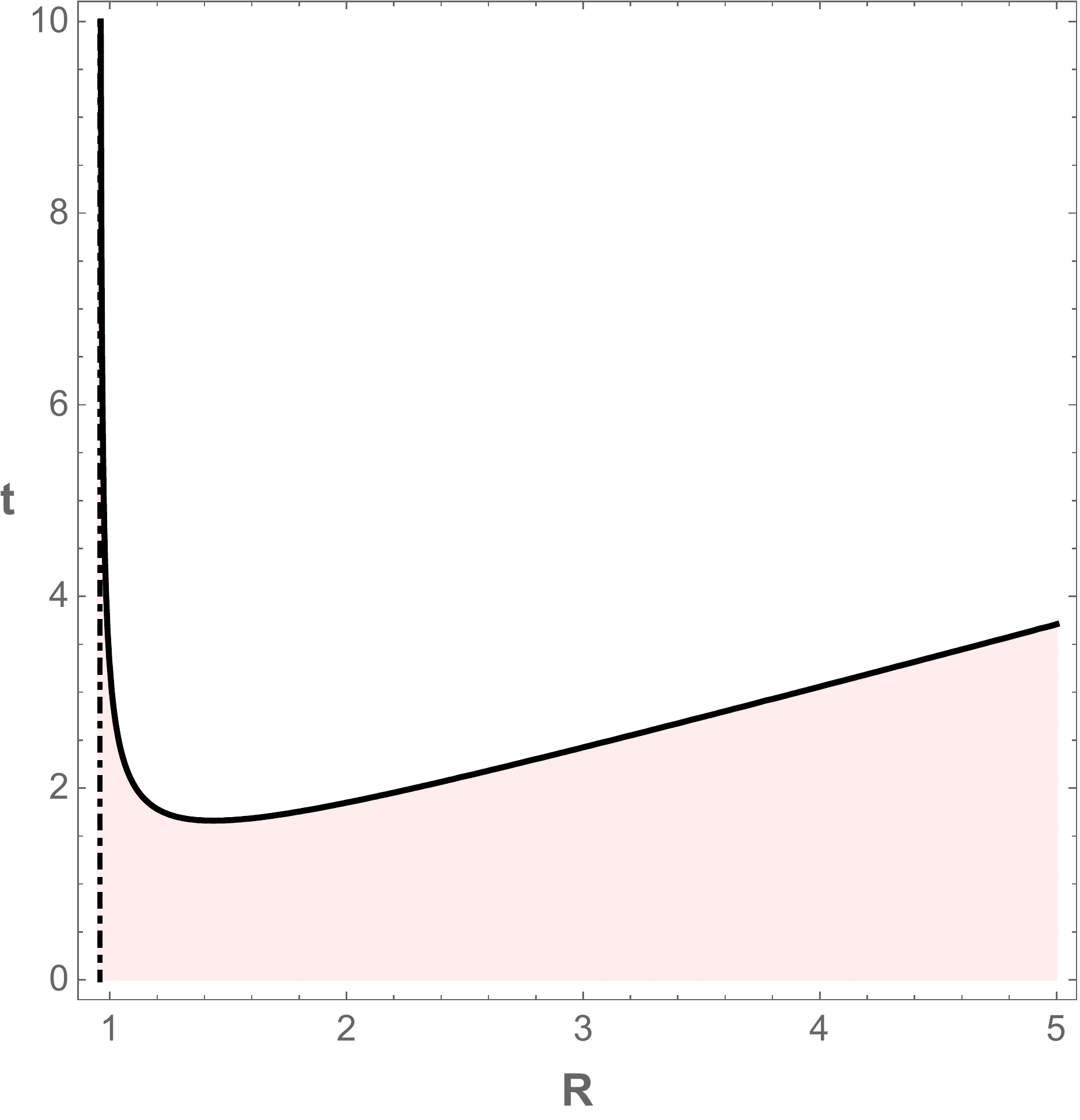}
\caption{\label{fig:1-1} The black line indicates the location of the trapping horizons in McVittie spacetime for a dust background model. The white zone corresponds to the regular region while the light pink corresponds to the anti-trapped region, respectively. The dot dashed line denotes the location of the singularity. Here, we fixed $m_0 = 0.479$ as an example.}
\end{figure}

We show in Figures \ref{fig:1} and \ref{fig:1-1} the location of the trapping horizons in the McVittie spacetime for the $\Lambda$CDM model \footnote{We are considering the minimal 6-parameter $\Lambda$CDM model; we assume that the curvature density parameter $\Omega_{k} =0$ and the equation of state parameter $w = -1$. The radiation density is neglected, $\Omega_{\rm rad} = 0$. Thus $\Omega_m + \Omega_\Lambda = 1$, being $\Lambda > 0$.}  and for a dust background. In the first case, the corresponding Hubble factor is $H(t) = H_0 \coth{(3/2 H_0 t)}$, where $H_0$ is the Hubble constant, while in the second case H(t) = 2/3t \cite{lak+11}. The essential difference between these two cosmological models is the asymptotic behaviour of the Hubble factor in the future: $H(t)\rightarrow H_0$ for $t\rightarrow \infty$ in the $\Lambda$CDM model whereas $H(t)\rightarrow 0$ for $t\rightarrow \infty$ in a dust-dominated background. 

There is a moment in time, denoted $t_{\star}$, when just one trapping horizon exists. It can be computed by solving the following equation:
\begin{equation}
m_0 H(t_{\star}) = \frac{1}{3 \sqrt{3}}.  
\end{equation}
If the cosmological background corresponds to the $\Lambda$CDM model,
\begin{equation}
{t_{\star}}_{\mathrm{\Lambda CDM}} =\frac{2}{3 H_0} \mathrm{arcoth}\left(\frac{1}{3\sqrt{3} m_0 H_0}\right),    
\end{equation}
and when the cosmological background is dust-dominated,
\begin{equation}
{t_{\star}}_{\mathrm{dust}} =2 \sqrt{3} m_0.
\end{equation}
Inspection of Figures \ref{fig:1} and \ref{fig:1-1} reveals that when $t > t_{\star}$ there are two trapping horizons: an inner (denoted $R_{-}$) and an outer one (denoted $R_{+}$), such that $R_{+}> R_{-}$. No trapping horizons are present for $t<t_{\star}$. In both figures, the regular region of the spacetime is indicated in white while the anti-trapped region is painted in light pink. The dot-dashed curve indicates the location of the singularity. 

\begin{figure}[t]
\includegraphics[width=8cm]{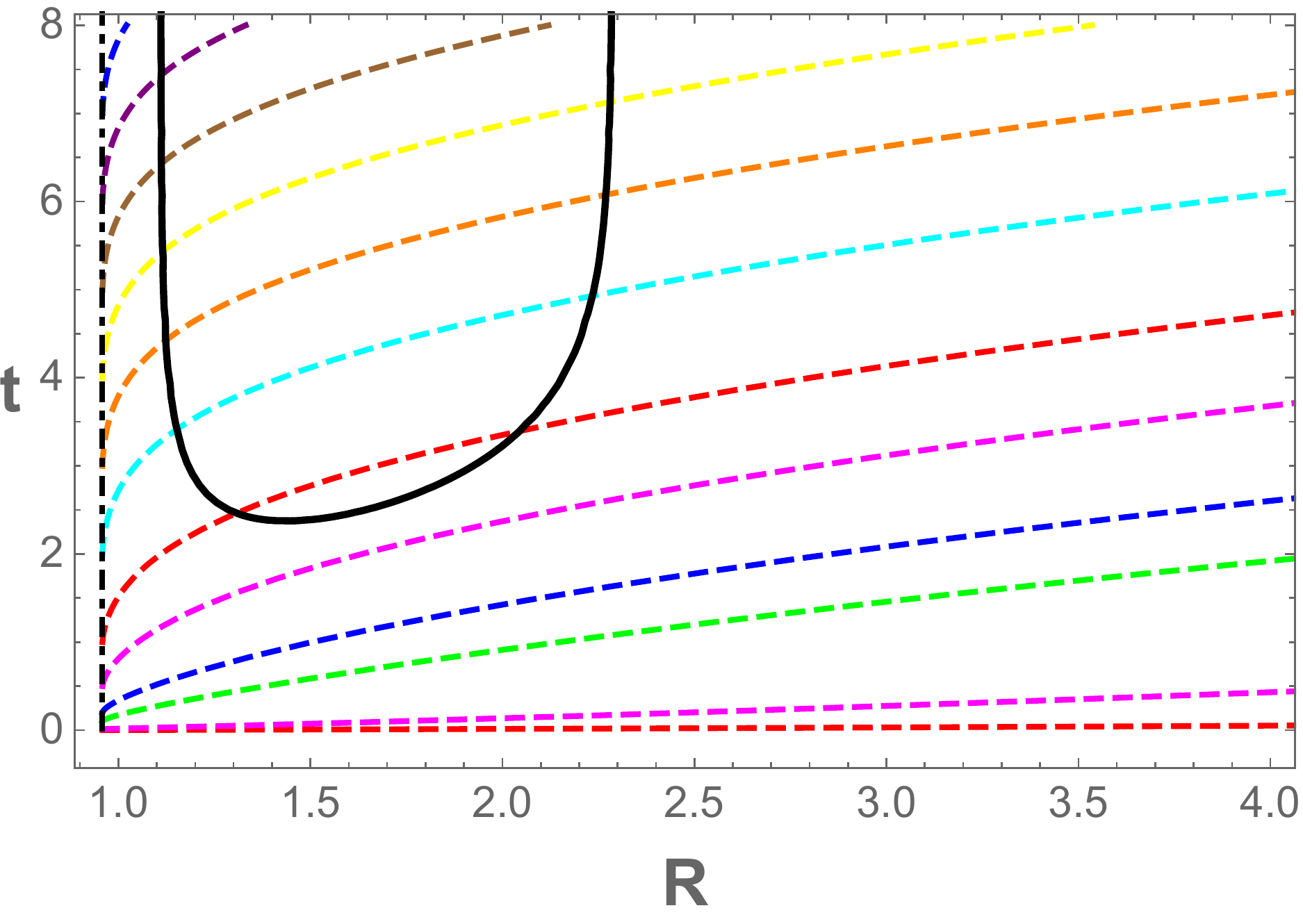}
\caption{\label{fig:2} Radial outgoing null geodesics in McVittie spacetime for the $\Lambda$CDM model. The black line indicates the location of the trapping horizons and the dot-dashed line denotes the singular surface $R = 2m_0$.  Here, we fixed $m_0 = 0.479$ and $H_0= 1/3$ as an example.}
\end{figure}

\begin{figure}[t]
\includegraphics[width=8cm]{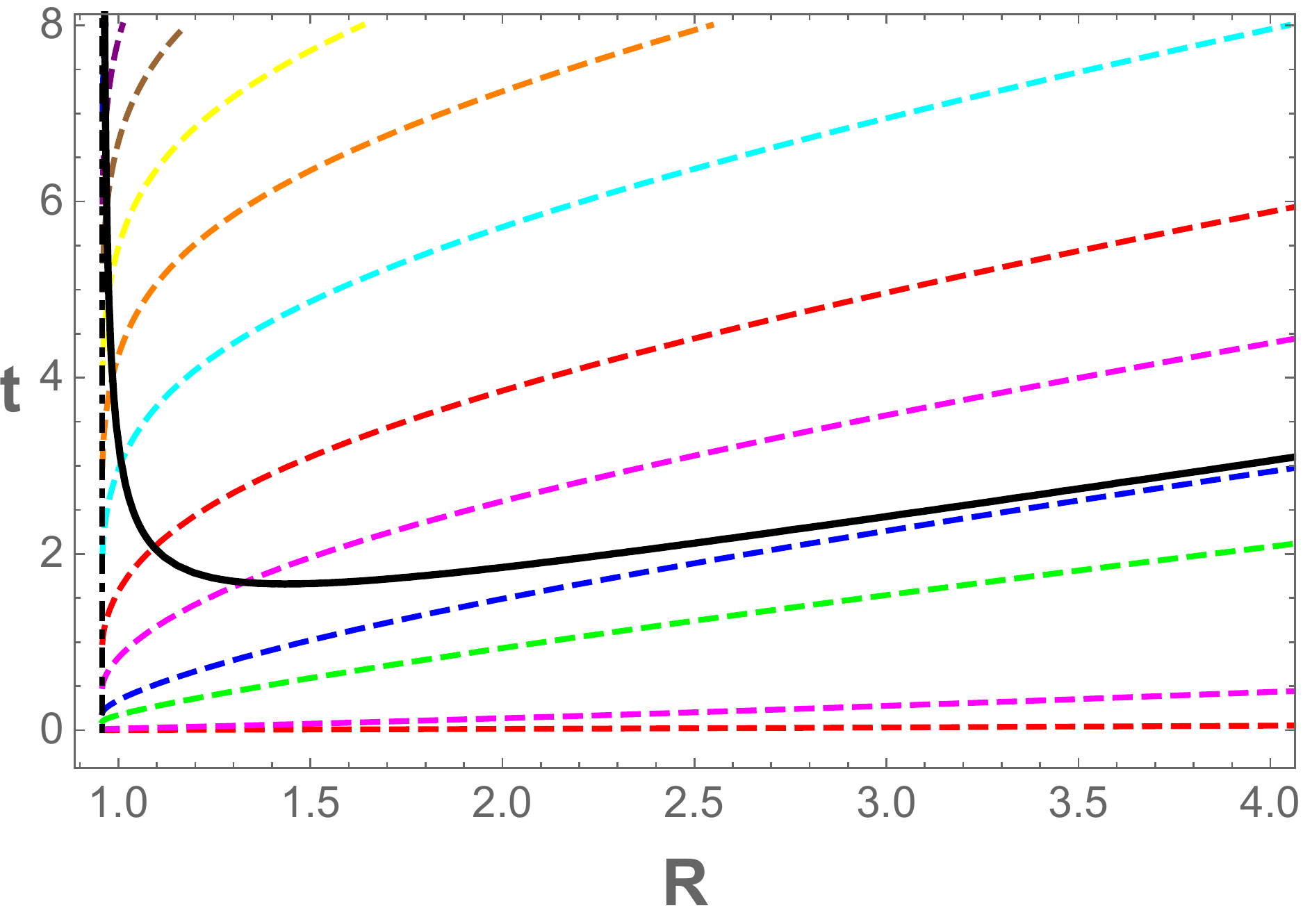}
\caption{\label{fig:2-2} Radial outgoing null geodesics in McVittie spacetime for a dust cosmological background. The black line indicates the location of the trapping horizons and the dot-dashed line denotes the singular surface $R = 2m_0$.  Here, we fixed $m_0 = 0.479$  as an example.}
\end{figure}

Due to spherical symmetry, the equation for the ingoing and outgoing radial geodesics can be derived by setting 
$d\theta = d\phi = 0$ in $ds^{2} = 0$, thus obtaining
\begin{equation}\label{out-in}
 \frac{dR}{dt} = \sqrt{1- 2m_0/R} \left(H R \pm \sqrt{1- 2m_0/R}\right),
\end{equation}
 where the ``$-$'' (``$+$'') corresponds to the ingoing (outgoing) case.  We see that $dR/dt > 0$ for the outgoing branch, i.e, such geodesics are always diverging. This result is consistent with the fact that $\theta_{\mathrm{out}} >0$, as mentioned above. We plot these trajectories in Figures \ref{fig:2} and \ref{fig:2-2}.

\begin{figure}[b]
\includegraphics[width=8cm]{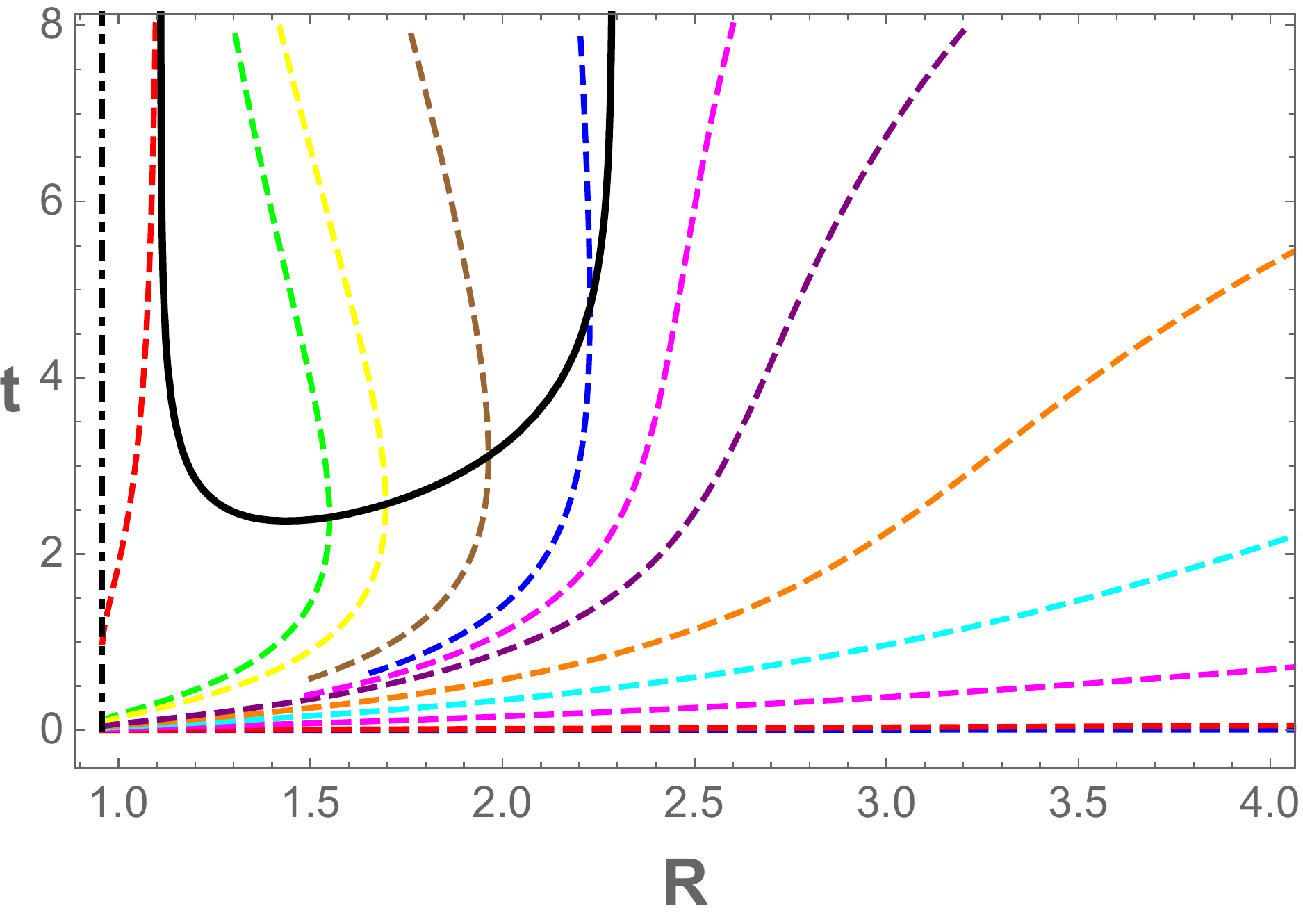}
\caption{\label{fig:3} Radial ingoing null geodesics in McVittie spacetime for the $\Lambda$CDM model. The black line indicates the location of the trapping horizons and the dot-dashed curve denotes the singular surface $R = 2m_0$. Here, we fixed $m_0 = 0.479$ and $H_0= 1/3$ as an example.}
\end{figure}

\begin{figure}[b]
\includegraphics[width=8cm]{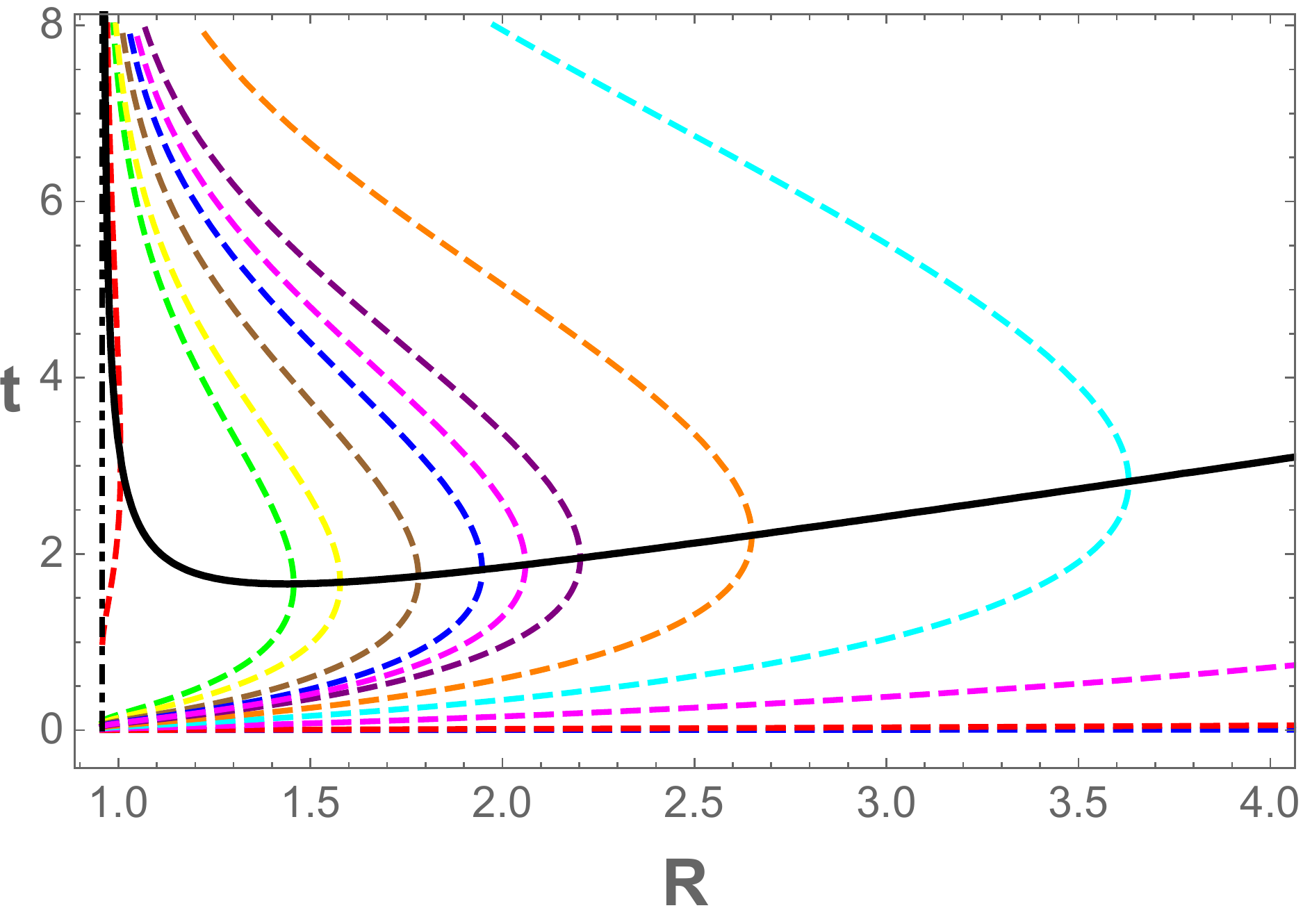}
\caption{\label{fig:3-3} Radial ingoing null geodesics in McVittie spacetime for a dust cosmological background. The black line indicates the location of the trapping horizons and the dot-dashed curve denotes the singular surface $R = 2m_0$. Here, we fixed $m_0 = 0.479$ as an example.}
\end{figure}

As seen from Eq. \eqref{out-in},  only radial ingoing null geodesics have a turning point, defined by $dR/dt = 0$, and specified by those values of the coordinates $R$ and $t$ that obey the following equation:
  \begin{equation}\label{trap-McVittie} 
   \tilde{f}(t,R) = H(t)^{2}R^{3} - R + 2m_0 =0.
  \end{equation}
  Clearly, Eq. \eqref{trap-McVittie} is equivalent to $\theta_{\mathrm{in}} = 0$. From Figures \ref{fig:3} and \ref{fig:3-3} we see that the radial ingoing geodesics are expanding in the anti-trapped region ($\theta_{\mathrm{in}} > 0$) and they converge in the regular zone ($\theta_{\mathrm{in}} < 0$). When they cross the trapping horizon ($\theta_{\mathrm{in}} = 0$ or equivalently $ \tilde{f}(t,R) = 0$), their convergence 
  changes sign.
   
 We emphasise that trapping horizons are not equivalent to event horizons in the context of dynamical spacetimes. As shown in Figures \ref{fig:2} and \ref{fig:2-2}, 
 some outgoing geodesics cross both $R_{-}$ and $R_{+}$. In Figure \ref{fig:3}, the green, yellow, brown, and blue geodesics cross $R_{+}$ and enter the regular region of the spacetime; as time goes by, they get closer and closer to the surface $R_{-}$. We stress that for finite values of the time coordinate, $R_{-}$ is just a trapping horizon. Only in the limit $t \rightarrow \infty$, $R_{-}$ becomes an event horizon and outgoing geodesics cannot travel out of the black hole. In fact, Kaloper and collaborators \cite{kal+10} proved that, under certain assumptions, the 
 analysis of the behavior of ingoing null geodesics in the limit $t \rightarrow \infty$ reveals the presence of an event horizon.

 
 Assuming that $H(t\rightarrow\infty)\rightarrow H_0$=constant, two surfaces are 
 particularly relevant
 in this spacetime \cite{kal+10}:
\begin{itemize}
\item A null surface at $R = R_{-}$, $t \rightarrow\infty$, where $R_{-}$ is the smaller positive root  of $\tilde{f} (t\rightarrow \infty,R) = H_{0}^2 R^3 - R + 2 m_0 = 0.$
\item A null surface at $R = R_{+}$, $t \rightarrow{\infty}$, where $R_{+}$ is the larger positive root  of $\tilde{f} (t \rightarrow \infty,R) = H_{0}^2 R^3 - R + 2 m_0 = 0.$
\end{itemize}  
%
If the null energy condition is satisfied, it was demonstrated in Ref. \cite{kal+10} that:
\begin{enumerate}
 \item Null ingoing geodesics in the regular region of the spacetime cross the surface $R = R_{-}$, $t \rightarrow{\infty}$ at a finite value of the affine parameter. 
 \item Such surface is regular (\emph{i.e.} all the squared curvature invariants 
 constructed with the Riemann tensor and its contractions 
 are finite on it)\footnote{
 Lake and Abdelqader \cite{lak+11} proved that, in the case $H_0 = 0$, $R = R_{-} = 2 m_0$ for $t \rightarrow \infty$ is also a regular surface in the sense defined above. }. 
 \item Once the geodesics transverse the surface $R = R_{-}$, $t = t_{\infty}$, they are in a trapped region since $\theta_{\mathrm{in}} \theta_{\mathrm{out}} > 0$ there. 
 \end{enumerate}
   Consequently, those geodesics that go through the surface $R = R_{-}$, $t \rightarrow {\infty}$, will not cross it again in the opposite direction. This is precisely the situation in the presence of an event horizon. The conclusion is that the solution represents a black hole with an horizon at $R = R_{-}$ at large times.
   
 Using a similar procedure as the one describe above, Lake and Abdelqader (see Appendix D in \cite{lak+11}) also demonstrated that in the case $H(t\rightarrow\infty)\rightarrow H_0$=0 (for instance, the dust-dominated background previously discussed), the McVittie metric represents a black hole in the future. 

Having presented the main features of the McVittie solution, we introduce in the next section the cosmological background model that we adopt in this work.
 
 \section{Scale factor for a bouncing cosmological model}
 
Cosmological models that display a bounce solve by construction the initial singularity problem, as well as the horizon and flatness problems of the standard cosmological model
\footnote{See Ref. \cite{Novello2008} for a review}. 
Such models can also 
produce primordial
cosmological perturbations from  vacuum fluctuations, with an almost scale-invariant spectrum
\cite{Peter:2008qz}, and can be viewed either as an
alternative or a complement  to inflation (see for instance Ref.
\cite{Falciano:2008gt}).
Typically, models with a bounce
join a contracting phase, in which the Universe was very large and almost flat initially,
to a subsequent expanding phase.
The bounce can be either generated classically (see e.g. Refs.
\cite{Wands:2008tv, Ijjas:2016tpn, Galkina:2019pir}), 
or by
quantum effects (see e.g. Refs.
\cite{Peter:2008qz, Almeida:2018xvj, Bacalhau:2017hja, Frion:2018oij}). Since our aim is to investigate the effects of the bounce on the McVittie solution, with no intention at this stage to build a complete cosmological model, our choice of the regular model will be guided by simplicity. 
The expression for the scale factor we adopt as background in our work, given by
 \begin{equation}
 a(t) = a_0 \left[1 +
\left( \frac{t}{t_0}\right )^{2}\right]^{1/3},  
 \end{equation}
was found in 
\cite{Celani:2016cwm}
by considering quantum corrections 
to the classical evolution of the scale factor. The corrections were obtained by
solving the Wheeler-deWitt equation in the presence of a single perfect fluid, 
in the framework of the 
de Broglie-Bohm quantum theory \cite{Pinto-Neto:2013toa}. 
Other quantization methods yield 
the same evolution for the scale factor, see Refs.
\cite{Ashtekhar2006, Taveras2008, Bergeron2014}. Notice that the scale factor reduces to that of dust for $t>>t_0$, and leads to an evolution that is dominated near the bounce by an effective fluid with negative energy density that scales as $a^{-6}$, as can be seen from Friedman's equation\footnote{ 
Limits on the parameters $a_0$ and $t_0$ can be found for instance in \cite{Frion2020} }.
\section{McVittie spacetime in a bouncing cosmological model}
 
Our goal is to compute and analyze the causal structure of the McVittie spacetime in a classical cosmological bouncing model.  As we discussed above, under certain assumptions, the McVittie metric represents a black hole. In what follows, we will examine how this solution behaves before, during, and after the bounce, and whether a black hole is present in any of these stages of cosmological evolution. 
 
\subsection{Trapping horizons and null geodesics} \label{th-nullgeo}
  
\begin{figure}[t]
\includegraphics[width=8cm]{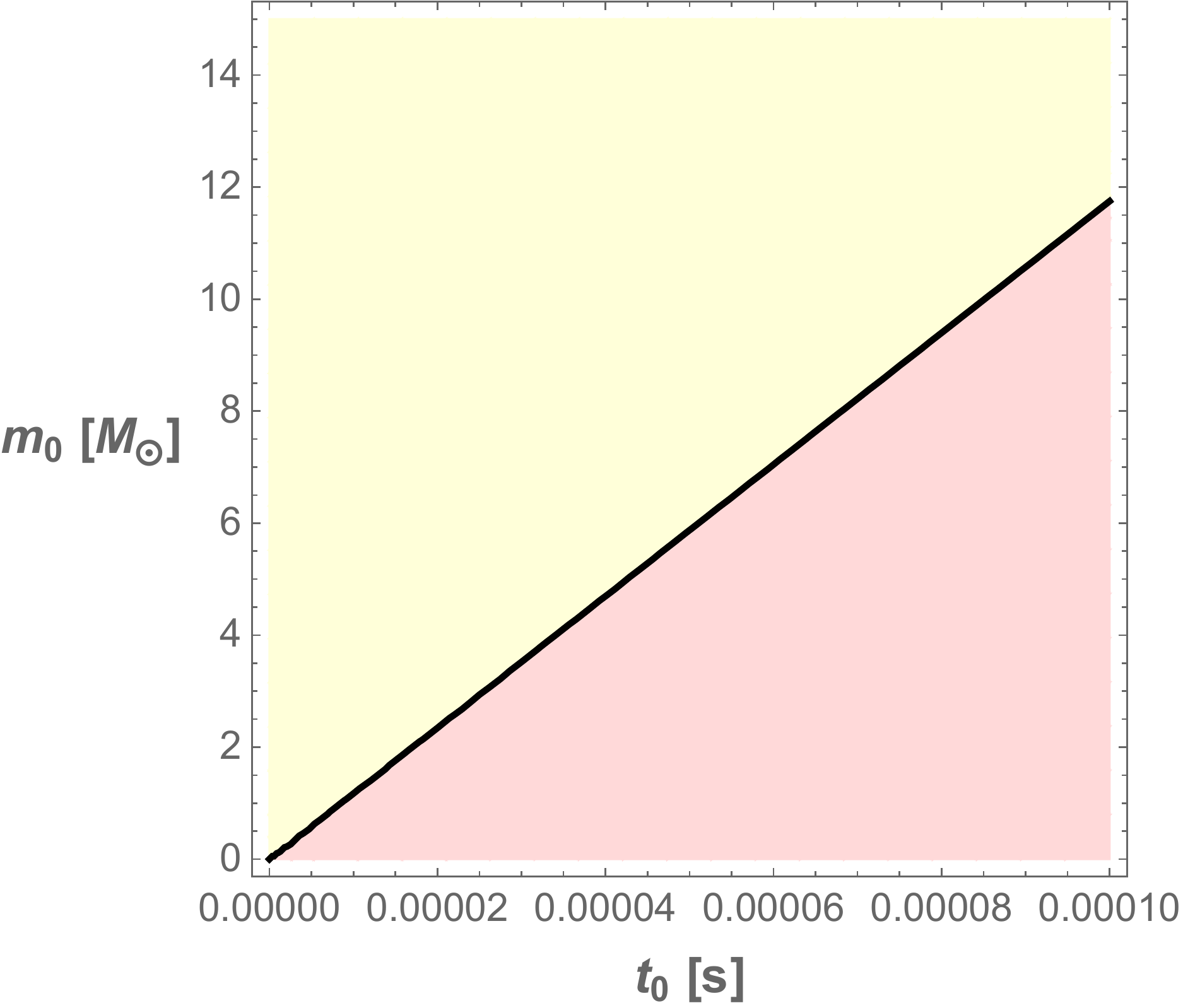}
\caption{\label{fig:7-7} The black line indicates the condition $ \sqrt{3} G m_0/c^3 = t_0$. The light yellow zone corresponds to values of $m_0$ and $t_0$ such that $ \sqrt{3} G m_0/c^3 > t_0$ while the light red region satisfy the condition $ \sqrt{3} G m_0/c^3 < t_0$. Here, $m_0$ is in units of  solar masses $M_{\odot}$.}
\end{figure}

We begin by computing the trapping horizons using Eq. \eqref{trap-McVittie} and the Hubble factor
\begin{equation}\label{hubble-factor}
 H(\tilde{t}) = \frac{2}{3 t_0}\frac{\tilde{t}}{1+\tilde{t}^{2}},  
\end{equation}
in the time interval $- \infty < \tilde{t} < + \infty$
\footnote{Notice that, contrary to the cases studied in \cite{lak+11}, $H$ goes to zero at $t=0$.}. In Eq. \eqref{hubble-factor}, the new variable $\tilde{t}$ is defined as $\tilde{t} \equiv t/t_0$. In what follows, to simplify the notation, we replace $\tilde{t}$ by $t$. We rewrite Eq. \eqref{trap-McVittie} recovering the corresponding units:
\begin{equation}
\frac{H(t)^{2}}{c^2}R^{3} - R + 2 \frac{G m_0}{c^2} =0.  
\end{equation}
Defining a new dimensionless variable as $x \equiv R/(G m_0/c^2)$, the latter equation takes the form
\begin{equation}\label{th}
  \alpha^{2} H(t)^{2} x^3 -x + 2 = 0,
\end{equation}
where $\alpha = G m_0 /c^3$. In order to solve this cubic equation, we express \eqref{th} as \cite{nic93}
\begin{equation}\label{th1}
  \alpha^{2} H(t)^{2} x^3 - 3 \alpha^2 H(t)^2 \delta^{2} x + 2 = 0.
\end{equation}
Here, $\delta^2 \equiv 1/(3 \alpha^2 H(t)^2)$. By performing the change of variables $x = 2 \delta \sin{\phi}$, and after some simple algebraic manipulations, Eq. \eqref{th1} becomes
\begin{eqnarray}
2 \alpha^2 H(t)^2 \delta^3 \left(4 \sin^3{\phi} - 3 \sin{\phi}\right) + 2 & = & 0, \nonumber \\
-    \alpha^2 H(t)^2 \delta^3 \sin{3\phi} + 1 & =& 0.\label{cond}
\end{eqnarray}
Trapping horizons exist only if $0 < \sin{3\phi} < 1$. In particular, let us focus on the case where the radial coordinate of the trapping horizons coincides; this occurs for $\sin{3\phi} = 1$, which in terms of Eq. \eqref{cond} takes the form
\begin{equation}\label{th2}
3\sqrt{3} \frac{G m_0}{c^3} H(t) =1, \Rightarrow t^2 - 2\sqrt{3} \frac{G m_0}{c^3 t_0} t + 1 = 0.
\end{equation}
The solutions for  \eqref{th2} are
\begin{equation}
 t = \sqrt{3}\frac{G m_0}{c^3 t_0} \pm \sqrt{\left(\sqrt{3}\frac{G m_0}{c^3 t_0}\right)^2 - 1}.
\end{equation}
Thus, depending on the value of the mass of the central source $m_0$ and $t_0$, which fixes the time scale of the bounce, we have the following three cases\footnote{Since we are now working with the radial dimensionless coordinate $x$, we denote the inner and outer radial coordinate of the trapping horizons as $X_{-} \equiv R_{-}/(G m_0/c^2)$ and  $X_{+} \equiv R_{+}/(G m_0/c^2)$.}:
\begin{enumerate}
\item If $\frac{\sqrt{3}  G m_0}{c^3} > t_0$, then there are two values of $t$ where $X_{-} = X_{+}$.
\item If $ \frac{\sqrt{3} G m_0}{c^3} = t_0$, then there is only one value of $t$ where $X_{-} = X_{+}$.
\item If $ \frac{\sqrt{3} G m_0}{c^3} < t_0$, then there is no value of $t$ where $X_{-} = X_{+}$.
\end{enumerate}
According to Frion and collaborators \cite{Frion2020}, $10^{3} \; t_{\mathrm{Planck}} < t_{0} < 10^{40} \; t_{\mathrm{Planck}}$, that is $t^{\mathrm{min}}_{0} = 10^{-41} \; \mathrm{s} < t_{0} < t^{\mathrm{max}}_{0} =10^{-4} \; \mathrm{s}$. We show in Figure \ref{fig:7-7} the range of allowed values for $m_0$ (in units of solar masses $M_{\odot}$) according to the three conditions described above: the black line indicates the condition $ \sqrt{3} G m_0/c^3 = t_0$. The light yellow zone corresponds to values of $m_0$ and $t_0$ such that $ \sqrt{3} G m_0/c^3 > t_0$, while the light red region satisfies the condition $ \sqrt{3} G m_0/c^3 < t_0$. 

From a physical point of view, it seems more attractive to consider a bounce time scale close to the allowed upper limit ($t^{\mathrm{max}}_{0} =10^{-4} \; \mathrm{s}$), that is, in the limit between a classical and quantum bounce. In such scenario, matter inhomogeneities that existed in the contracting phase might be able to go through the bounce and also be present in the expanding epoch. On the other hand, Carr and K\"{u}hnel \cite{car+20} have recently showed that primordial black holes in the mass range $10 \; M_{\odot} < M < 10^2 \; M_{\odot}$ could be relevant to provide a fraction of the dark matter in the universe as well to explain the observed LIGO/Virgo coalescence events in the mass range $\mathcal{O}(10)\; M_{\odot}$. Therefore, in what follows we choose for $m_0$ and $t_0$, $m_0 = 50 M_{\odot}$ and $t_0 = 5 \times 10^{-5}$ s. The structure of the trapping horizons that will be analysed in this work corresponds to case 1.

\begin{figure}[t]
\includegraphics[width=8cm]{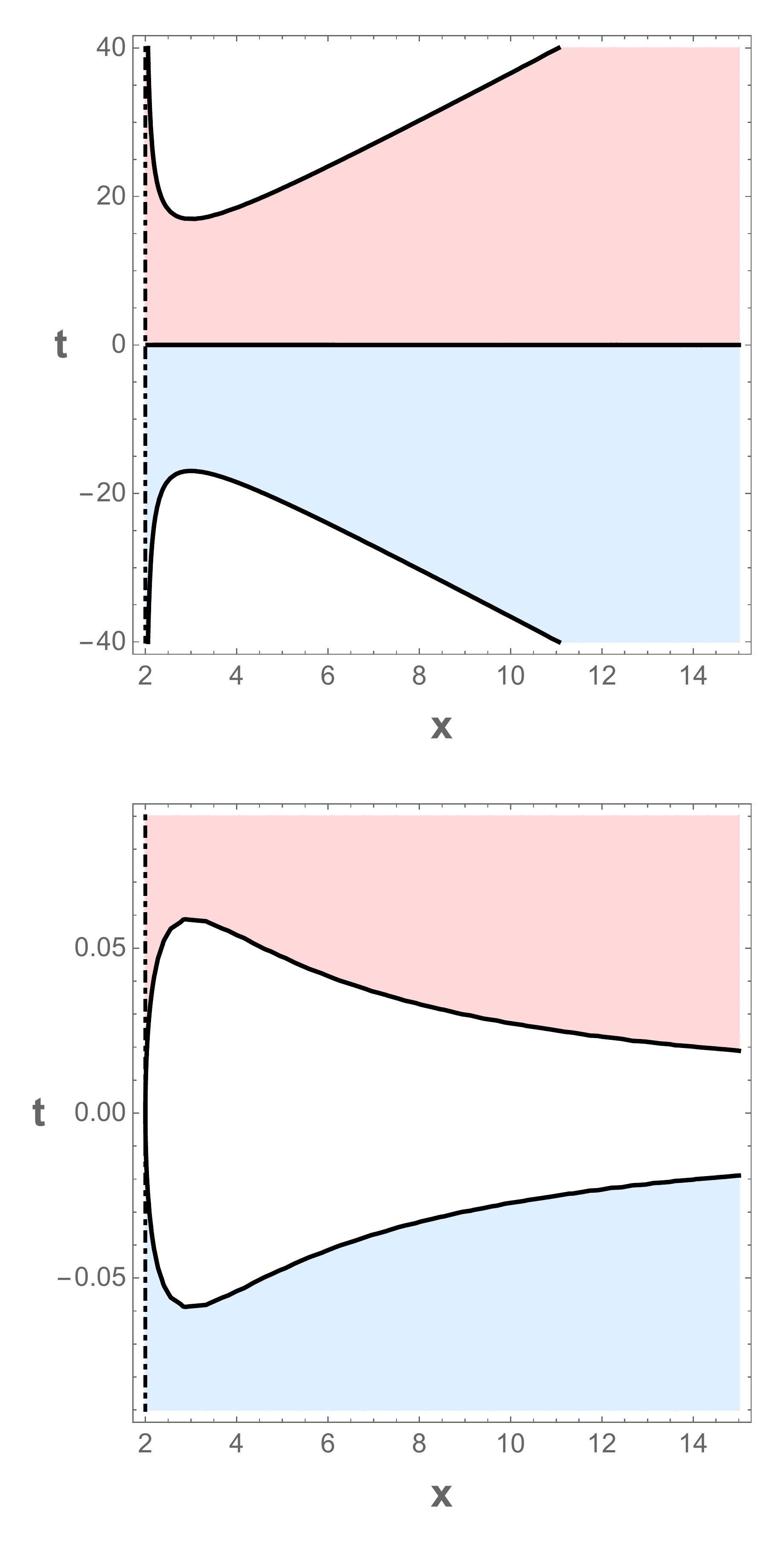}
\caption{\label{fig:4} The black lines indicate the location of the trapping horizons. The white zones correspond the regular regions, the light pink zone indicates the anti-trapped region and the light blue zone denotes the trapped region. The dot dashed curve denotes the location of the singularity. The lower plot zooms into the region near  the bounce. Here, $m_{0} = 50 \; M_{\odot}$, and $t_{0} = 5 \times 10^{-5}$ s}.
\end{figure}

 The trapping horizons are plotted  in Figure \ref{fig:4}: the lower plot zooms into the region near  the bounce. Very close to $t =0$, just before and after the bounce, there is a trapping horizon. This surface is absent, for instance, in McVittie spacetime for the dust-dominated background (see Figure \ref{fig:1-1}). In Section \ref{sec:discussion}, we argue that this horizon is a particular feature related to the presence of the bounce.

There are two additional trapping horizons for negative and positive values of the cosmic time, respectively.  The shape of these surfaces is qualitatively similar compared to the trapping horizons in the McVittie solution for the dust background (see Figure \ref{fig:1-1}): for $t > 0$ there is a moment in time when a single trapping horizon begins to exist and immediately after an inner $X_{-}$ and outer $X_{+}$ trapping horizons emerge. In the limit $t \rightarrow \infty$, $X_{-} \rightarrow 2 $. 

The second root of Eq. \eqref{trap-McVittie}, namely $X_+$, is given by \cite{far15}
\begin{equation}
X_{+} = \frac{c^3}{G m_0}\left[\frac{1}{H(t)}\cos{\psi(t)} - \frac{1}{\sqrt{3}H(t)} \sin{\psi(t)}\right]
 \end{equation}
where $\sin{3 \psi(t)} = 3 \sqrt{3}G m_0 H(t)/c^3$. For the Hubble factor given by Eq. \eqref{hubble-factor}, $X_+ \rightarrow \infty$ when $t \rightarrow \pm \infty$, and thus $X_+$ becomes a FLRW null infinity \cite{kal+10}.
  
The shape of the curve of the horizons is symmetric with respect to the axis $t = 0$. In the limit $t \rightarrow - \infty$, $X_{-} \rightarrow 2 $. As we approach the bounce, $X_{-}$ increases while $X_{+}$ decreases up to they merge.  The symmetry for the trapping horizons with respect to $t =0$ is rooted in the equation that defines these surfaces (see Eq. \eqref{trap-McVittie} or \eqref{th}), which is quadratic in the Hubble factor. 
 
 The white zones in Figure \ref{fig:4} indicate the regular regions of the spacetime ($\theta_{\mathrm{in}} \theta_{\mathrm{out}} < 0$), the light pink zone corresponds to the anti-trapped region ($\theta_{\mathrm{in}} \theta_{\mathrm{out}} > 0$, being $\theta_{\mathrm{in}} > 0$ and $\theta_{\mathrm{out}} > 0$), and the light blue zone marks the trapped region ($\theta_{\mathrm{in}} \theta_{\mathrm{out}} > 0$, being $\theta_{\mathrm{in}} < 0$ and $\theta_{\mathrm{out}} < 0$). The dot-dashed curve marks the location of the singularity $x = 2 $ for $t$ finite.


\begin{figure}[t]
\includegraphics[width=8cm]{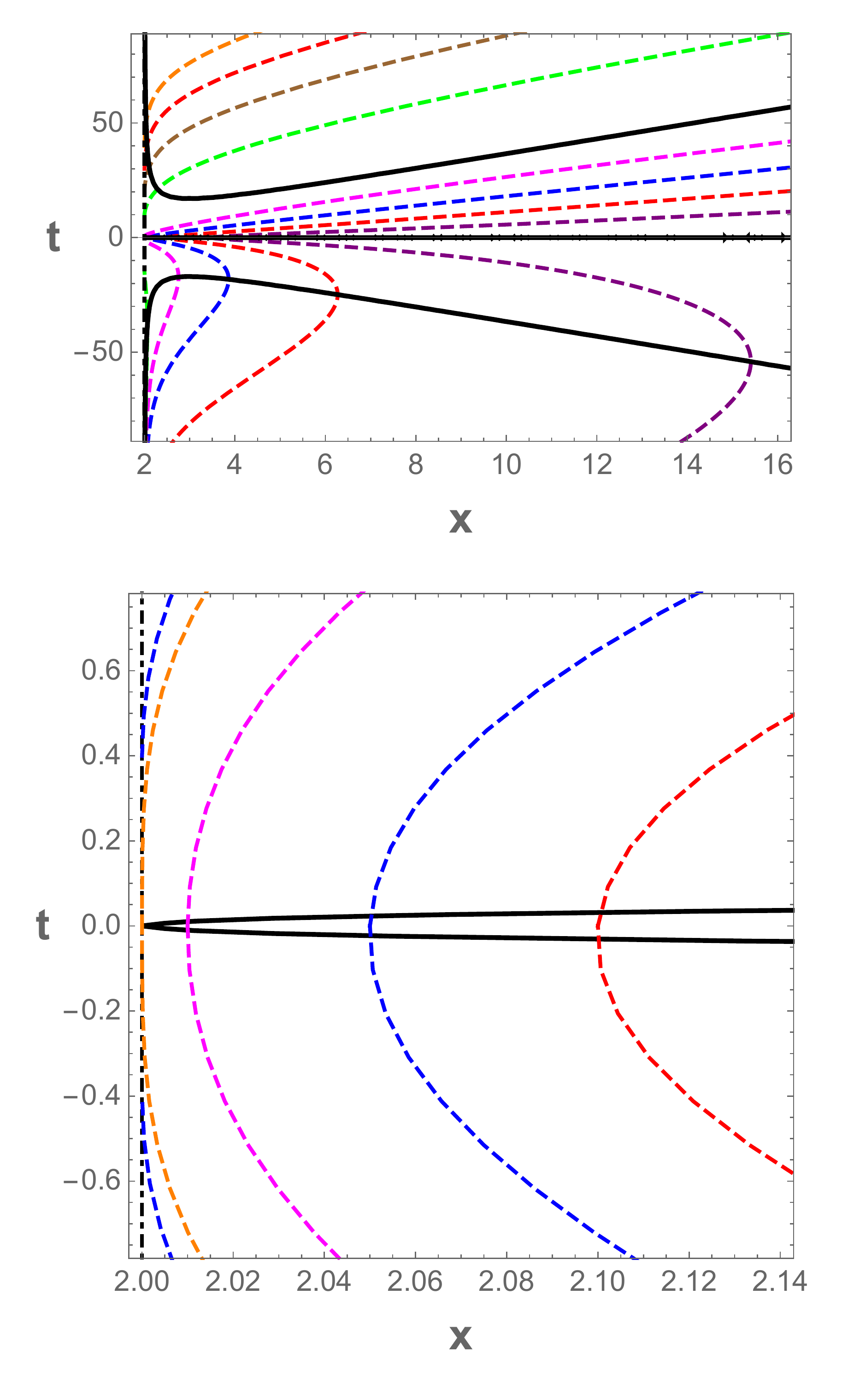}
\caption{\label{fig:5} Radial outgoing geodesics in  McVittie spacetime for a bouncing cosmological background. The black line indicates the location of the trapping horizons while the dot dashed curve denotes the location of the singularity. The lower plot zooms into the region near  the bounce. Here, $m_{0} = 50 \; M_{\odot}$, and $t_{0} = 5 \times 10^{-5}$ s.}
\end{figure}

\begin{figure}[t]
\includegraphics[width=8cm]{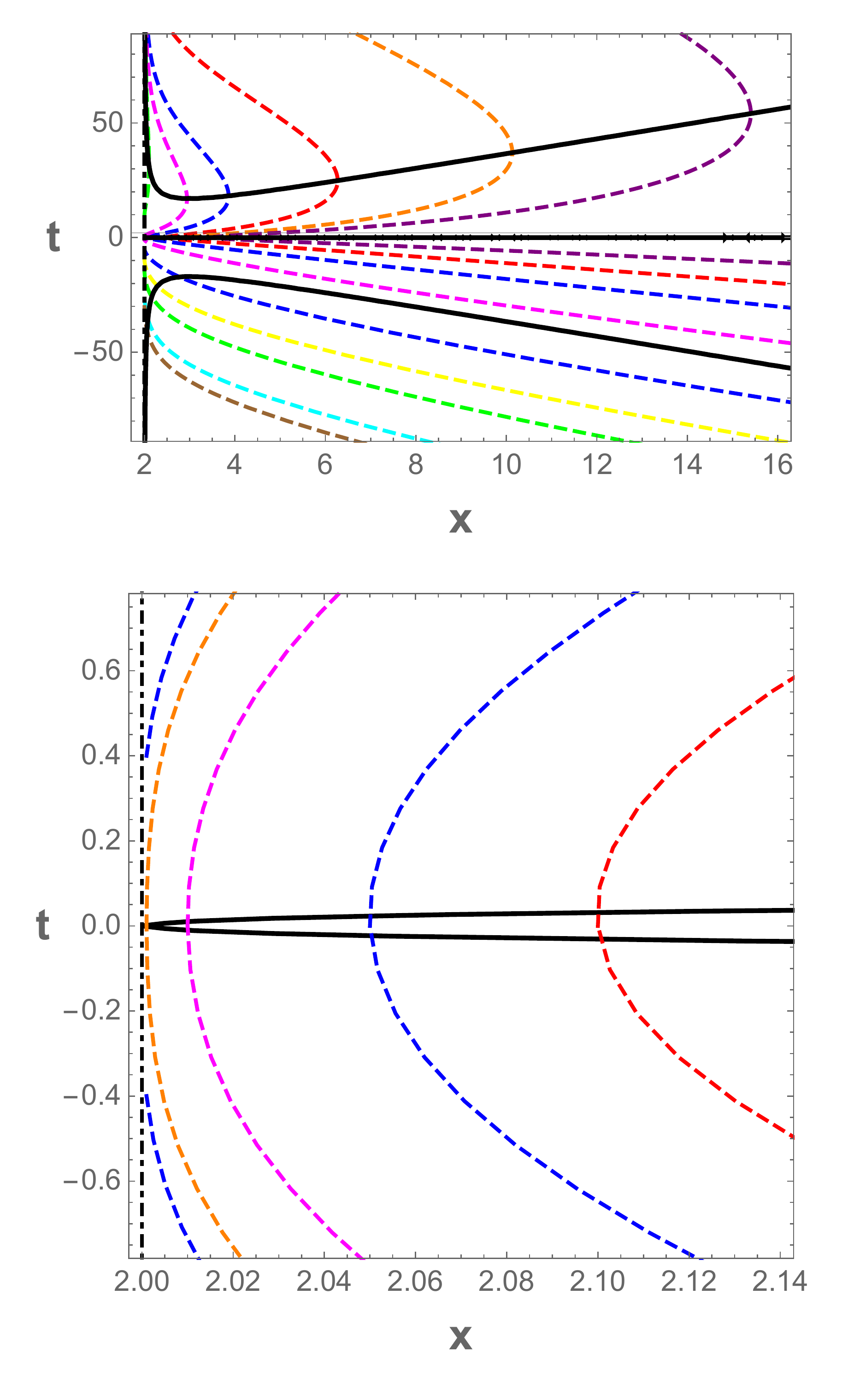}
\caption{\label{fig:6} Radial ingoing geodesics in  McVittie spacetime for a bouncing cosmological background. The black line indicates the location of the trapping horizons while the dot dashed curve denotes the location of the singularity. The lower plot zooms into the region near  the bounce. Here, $m_{0} = 50 \; M_{\odot}$, and $t_{0} = 5 \times 10^{-5}$ s.}
\end{figure}

We also compute the trajectories of ingoing and outgoing radial null geodesics by integrating Eq. \eqref{out-in}. The behaviour of the geodesics changes before and after the bounce:
\begin{itemize}
\item Outgoing null geodesics are always expanding ($(dx/dt)_{\mathrm{out}} > 0$) for $t > 0$, as can be  seen in Figure \ref{fig:5}. Radial ingoing geodesics expand in the anti-trapped region until they cross the trapping horizon; once in the regular region of the spacetime ${(dx/dt)}_{\mathrm{in}} < 0$, they all seem to tend asymptotically to the surface  $X_{-} = 2, \; t = \infty$, as shown in Figure \ref{fig:6}. Below, we will provide a more detail  analysis of the properties of the ingoing null geodesics in the limit $t \rightarrow \infty$, which is essential to establish whether a black hole is present.
\item Ingoing and outgoing radial null geodesics reverse their character for $t < 0$. Now, all ingoing trajectories are expanding to the past of the bounce (for increasing negative values of the $t$ coordinate). This is  in Figure \ref{fig:6}. Outgoing null geodesics converge in the trapped region. To the past of these geodesics, they seem to come from the surface $X_{-} = 2 $, $t = - \infty$ (see Fig. \ref{fig:5}). In Section \ref{before-bounce} we offer a possible interpretation of the McVittie solution before the bounce takes place.
\end{itemize}

\begin{figure}[t]
\includegraphics[width=8cm]{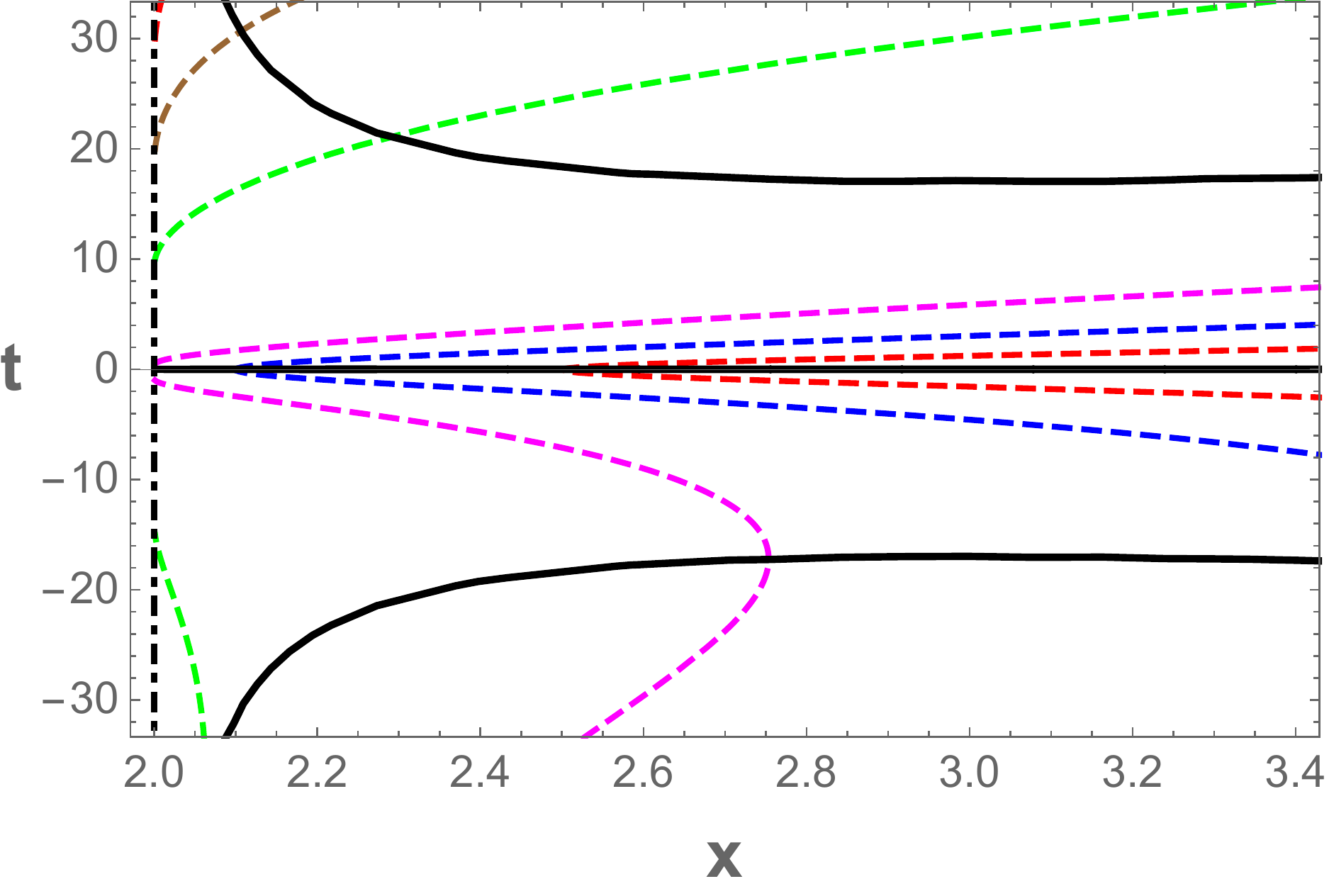}
\caption{\label{fig:11}Radial outgoing geodesics in  McVittie spacetime for a bouncing cosmological background in the region close to the bounce. The black line indicates the location of the trapping horizons while the dot dashed curve denotes the location of the singularity. Here, $m_{0} = 50 \; M_{\odot}$, and $t_{0} = 5 \times 10^{-5}$ s.}
\end{figure}

\begin{figure}[t]
\includegraphics[width=8cm]{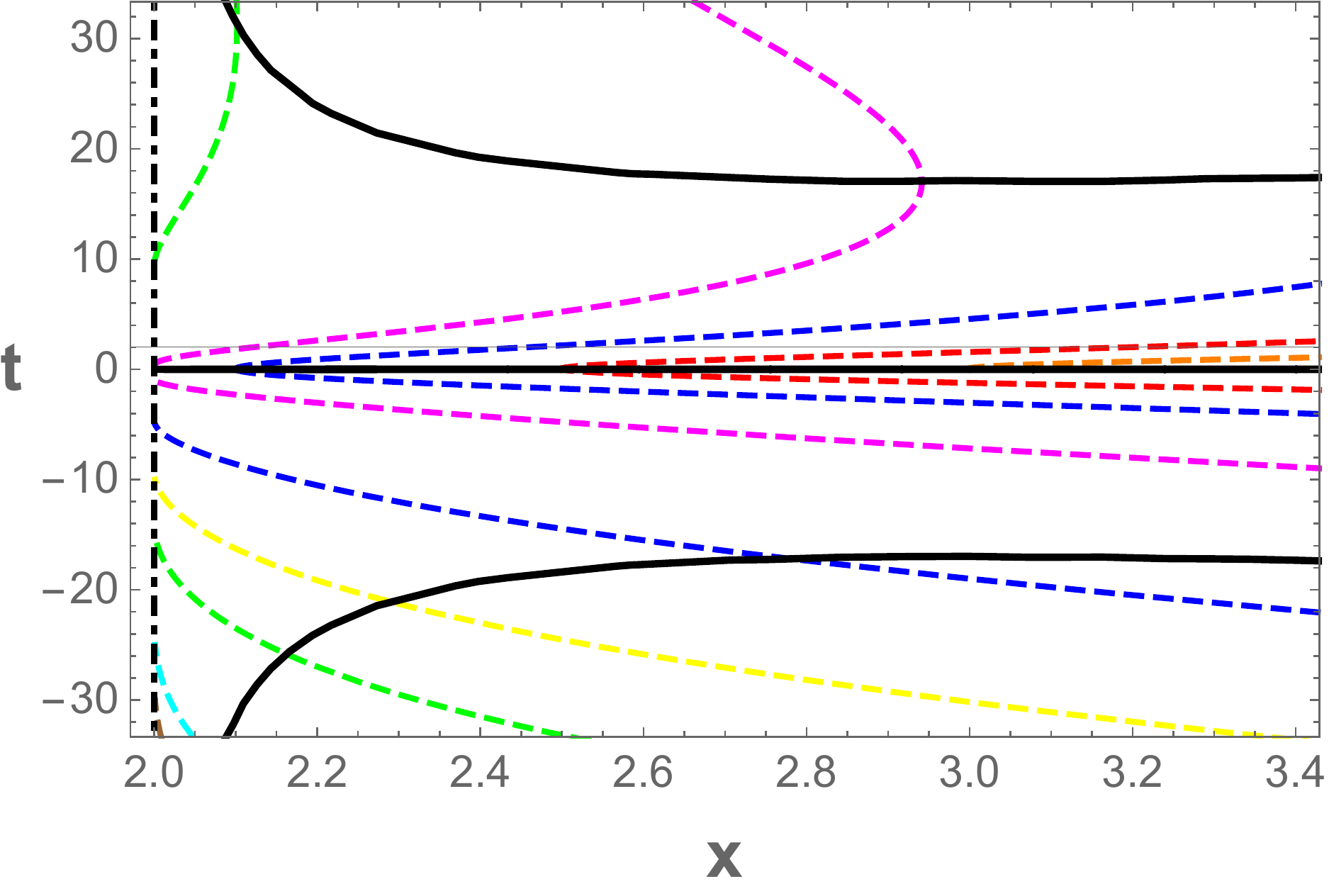}
\caption{\label{fig:12}Radial ingoing geodesics in  McVittie spacetime for a bouncing cosmological background in the region close to the bounce. The black line indicates the location of the trapping horizons while the dot dashed curve denotes the location of the singularity. Here, $m_{0} = 50 \; M_{\odot}$, and $t_{0} = 5 \times 10^{-5}$ s.}
\end{figure}

In Figures \ref{fig:11} and \ref{fig:12} we offer a close up of the region near the bounce showing the behaviour of the outgoing and ingoing null geodesics. Some null geodesics (painted in blue and red in Figure \ref{fig:11}) start at the remote past near $x = 2 $, go through the bounce and expand getting away from the central inhomogeneity. The same occurs for some ingoing geodesics. This does not happen in any McVittie model without a bounce analyzed so far in the literature; in those spacetimes the singularity at $x = 2m_0$, $t$ finite, lies in the causal past of all events, and thus it is regarded as a cosmological big bang singularity \cite{kal+10}. In the present model, the existence of a cosmological bounce renders only some spacetime trajectories geodesically incomplete.

The equation for the ingoing radial geodesics in terms of the Hubble factor \eqref{hubble-factor} takes the form
\begin{equation}\label{ingoing}
 \left.\frac{dx}{dt} \right\vert_{\mathrm{in}} =\frac{1} {\alpha}\sqrt{1- 2/x} \left( \frac{2 \; \alpha}{3}\frac{t}{1+t^{2}} x -  \sqrt{1- 2/x}\right),
\end{equation}
where  $\alpha = G m_0/(c^3 t_0)$. Before the bounce, the Hubble factor can be rewritten as $H(t) = -2 t/(3 t_0 (1+t^2))$ where $t >0$. Replacing into \eqref{ingoing}:
\begin{eqnarray}
\left.\frac{dx}{dt} \right\vert_{\mathrm{in}} & = & - \left[ \frac{1}{\alpha} \sqrt{1- 2/x} \left( \frac{2 \; \alpha}{3 }\frac{t}{1+t^{2}} x + \sqrt{1- 2/x}\right)\right] \nonumber \\
& = &- \left.\frac{dx}{dt} \right\vert_{\mathrm{out}},
 \end{eqnarray}
for $t >0$.  Thus, we can see that the trajectories of ingoing null geodesics, before the bounce, are the reflection of the trajectories of outgoing null geodesics after the bounce.

In the same way, we express the equation for the outgoing null geodesics before the bounce as:
\begin{eqnarray}
\left.\frac{dx}{dt} \right\vert_{\mathrm{out}} & = & \frac{1}{\alpha} \sqrt{1- 2/x} \left(- \frac{2 \; \alpha}{3}\frac{t}{1+t^{2}} x +  \sqrt{1- 2/x}\right)\nonumber \\
& = &  - \frac{1}{\alpha} \left[ \sqrt{1- 2/x}  \left( \frac{2 \; \alpha}{3 }\frac{t}{1+t^{2}} x -  \sqrt{1- 2/x}\right) \right]\nonumber\\
& = & - \left.\frac{dx}{dt} \right\vert_{\mathrm{in}},
\end{eqnarray} 
for $t > 0$. We conclude that the congruence properties of outgoing null geodesics before the bounce are the same that for ingoing null geodesics after the bounce.





\section{Spacetime structure after the bounce}

In order to prove that the surface $X_{-} = 2 $, $t = \infty$ is an event horizon the following conditions must be fulfilled: a) null ingoing radial geodesics reach $X_{-} = 2 $, $t = \infty$ in a finite interval of an affine parameter; b) the surface $X_{-} = 2 $, $t = \infty$ is regular.

To check if condition (a) is met, we make the following change of variables \cite{lak+11}:
\begin{eqnarray}
z & = & \sqrt{1 - \frac{2}{x}}, \label{z-l} \\
l & = & \frac{1}{1 + \alpha H(t)},\label{l-t}
\end{eqnarray}
where $0 \le z \le 1$ ($ 2 \le x < \infty$) and $0 < l < 1$. 

Since we are interested in the behavior of the geodesics for large values of $t$, we restrict the integration in terms of $l$ in the interval $0.75 \le l < 1$.

Given the expressions \eqref{z-l} and \eqref{l-t}, the equations for the ingoing geodesics \eqref{out-in} take the form
\begin{eqnarray}
\frac{dz}{dl}  & = &  \frac{\left(1-z^2\right)^2}{4 } \left[\frac{2 (1-l)}{l \left(1-z^2\right)} -z\right] \frac{dt}{dl}, \label{dif-z-l} \\
\frac{dl}{dt}  & = &  \frac{9 + l \left[ -\beta + \alpha \sqrt{-9 + \beta l}\right]}{3 \; l^4},\\
\beta & = & 18 + \left(\alpha^2 -9\right) l.
\end{eqnarray}

\begin{figure}[H]
\includegraphics[width=8cm]{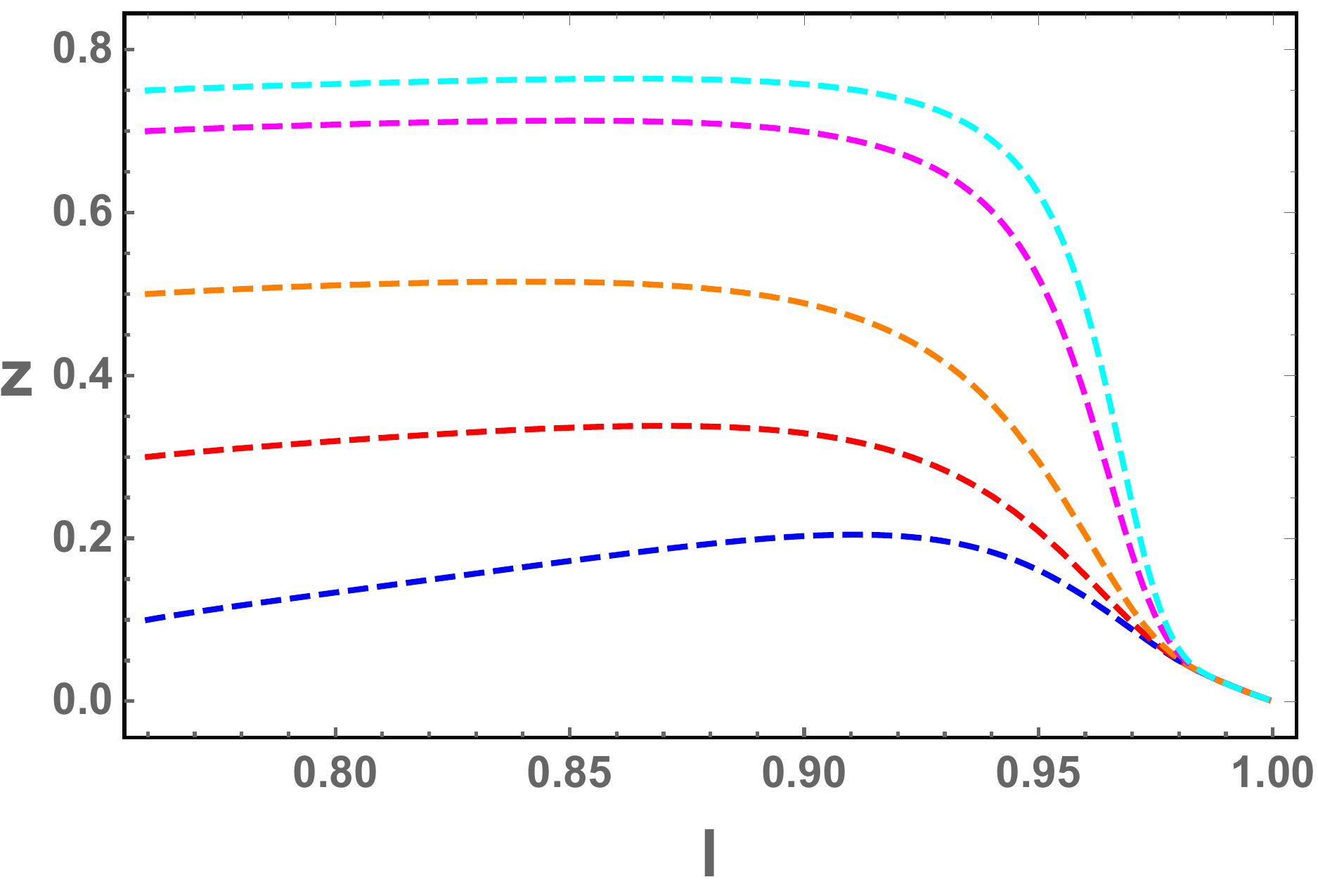}
\caption{\label{fig:16} Trajectories of radial ingoing null geodesics in the $z-l$ plane for McVittie spacetime in a bouncing cosmological model. Here, $m_{0} = 50 \; M_{\odot}$, and $t_{0} = 5 \times 10^{-5}$ s.}
\end{figure}

In Figure \ref{fig:16}, we plot the solution of the numerical integration of Eq. \eqref{dif-z-l} for five different initial conditions. Clearly, radial ingoing null geodesics reach the surface $x = 2 $ ($z = 0$) for $l = 1$.
\begin{figure}[H]
\includegraphics[width=8cm]{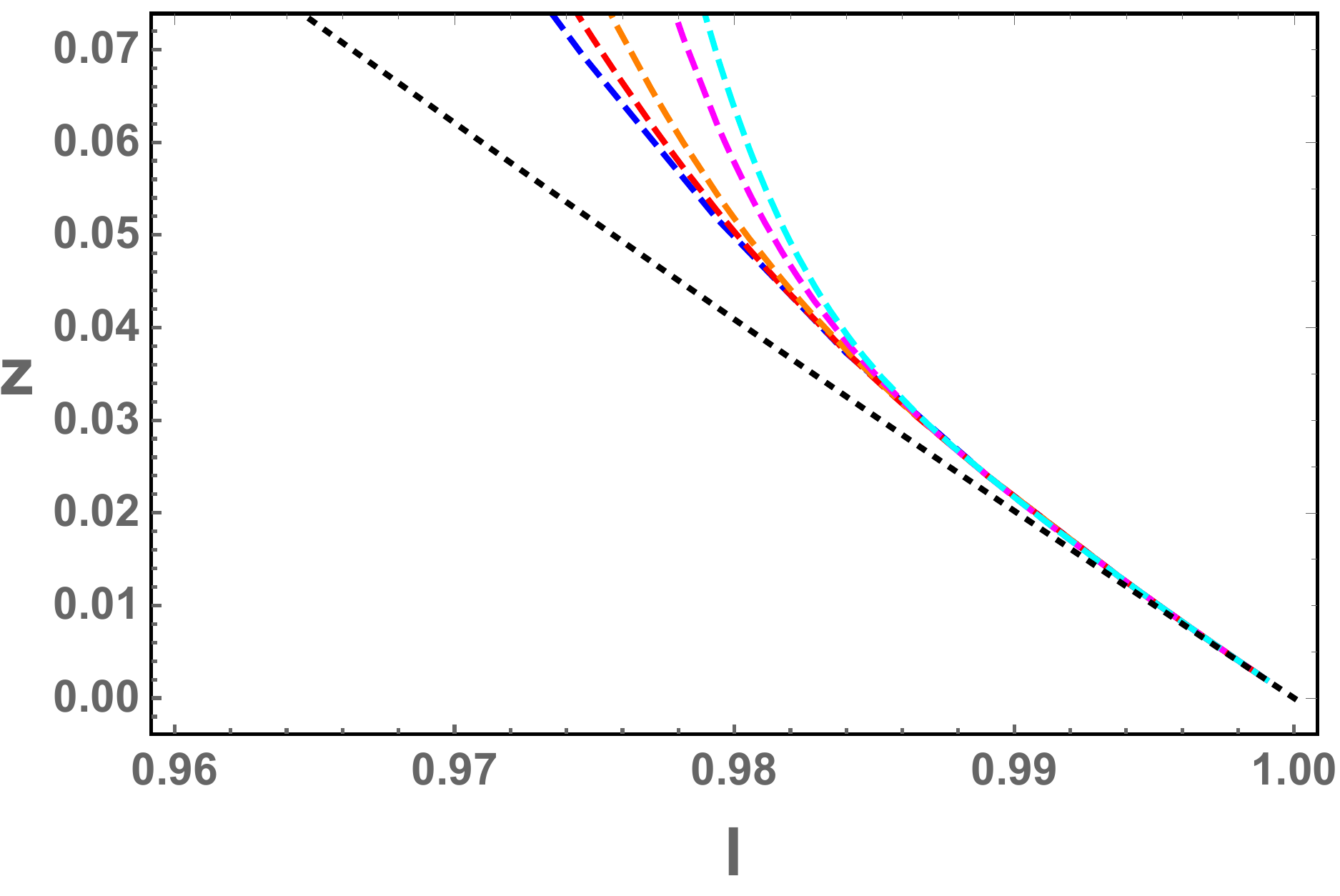}
\caption{\label{fig:17}Closeup in the $z-l$ plane showing the trajectories of radial ingoing null geodesics approaching the surface $x = 2$. Expression \eqref{aprox} is represented by the dotted black curve. Here,$m_{0} = 50 \; M_{\odot}$, and $t_{0} = 5 \times 10^{-5}$ s.}
\end{figure}

One way to decide whether condition (b), i.e. that $X_{-}$ is regular at large values of $t$, is valid or not is to calculate the components of the Riemann tensor 
using an appropriate \emph{Vierbein}, and evaluate them on null geodesics approaching the surface. Since the model given by Eq.\eqref{hubble-factor} behaves as a
model dominated by dust for large $t$, namely
\begin{equation} 
H(t) \approx\frac{2}{3 \;t},
\end{equation}
we can use the result (valid for large values of $t$) obtained in \cite{kal+10}:
 \begin{equation}
 \label{ngk}
t = \frac{2 x}{3\sqrt{1 - \frac{2 }{x}}}, \end{equation}
or, in terms of the variables $z$ and $l$ :
 \begin{equation}\label{aprox}
l z^3 - l z - 2 l + 2 =0.     
 \end{equation}
 Figure \ref{fig:17}
 shows that 
the asymptotic form 
 of the null ingoing geodesics (given by Eq.(\ref{aprox}))
 describes extremely well the result of the exact numerical integration.
%
Using GRTENSOR II \footnote{http://grtensor.phy.queensu.ca/}
, we have calculated the components of the Riemann tensor in the following \emph{Vierbein}:
\begin{equation}
    [e^j_{\:\mu}]=
    \begin{pmatrix}
\sqrt{H(t)^2r^2+f} & 0 & 0 & 0 \\
H(t)r & \frac{1}{\sqrt{1-\frac{2 m_0}{r}}} & 0 & 0\\
0 & 0 & r & 0 \\
0 & 0 & 0 & r\sin\theta
\end{pmatrix}
\end{equation}
The evaluation of the components on the null ingoing geodesics using \eqref{ngk} leads to finite values. Hence, the surface is regular. 
We have also evaluated, in the same limit, 
invariants built with first or second derivatives of tensors associated with curvature, such as $(\nabla_\mu R)(\nabla^\mu R)$ and $(\nabla_\mu \nabla_\nu R)(\nabla^\mu \nabla^\nu R)$, using RGTC \footnote{http://www.inp.demokritos.gr/∼sbonano/RGTC/}. While invariants of the first kind yield a finite result, those of the second kind are divergent, in agreement with the result found in \cite{kal+10}. Such a divergence does not influence the finiteness of the tidal forces, and is probably a consequence of the finite differentiability of the quantities describing the fluid \cite{Musgrave1995}.

We conclude that the solution contains a black hole after the bounce.

\begin{figure}[H]
\includegraphics[width=8cm]{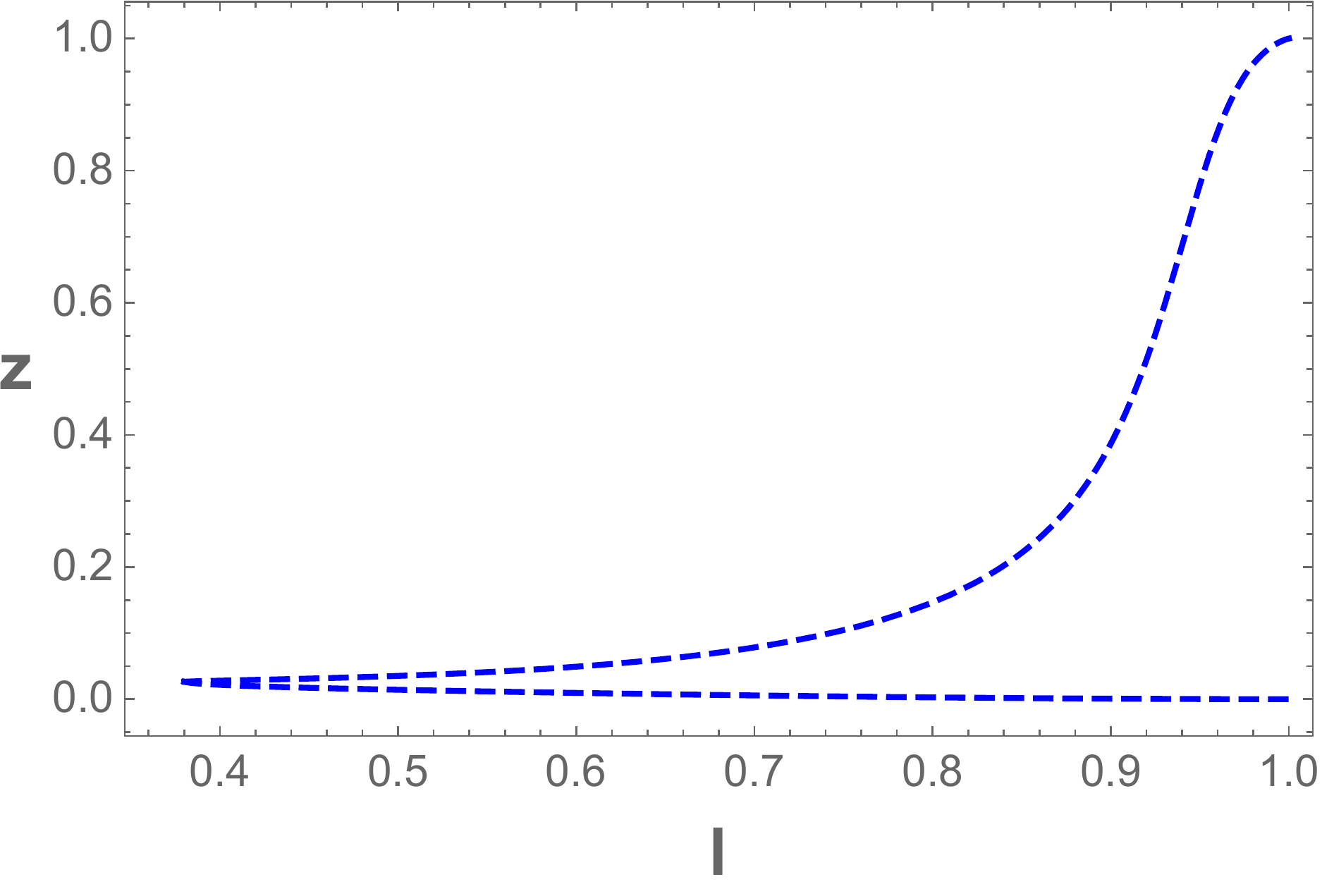}
\caption{\label{fig:17-1} Trajectory of a future-directed outgoing null geodesic that emerges from the singular point $(l,z) = (0,0)$ and reaches $z = 1$ ($x \rightarrow \infty$) in a finite interval of time.}
\end{figure}

Using the change of variables given by Eqs. \eqref{z-l} and \eqref{l-t}, we also integrate the trajectories of outgoing null geodesics. We show in Figure \ref{fig:17-1} the trajectory of a future-directed outgoing null geodesic that emerges from the singular point $(l,z) = (0,0)$, that is $(t,x) = (0,2)$ and reaches $z = 1$ ($x \rightarrow \infty$) in a finite interval of time. Thus, we see that a naked singularity is present in the expanding phase up to the formation of the black hole.


\section{Spacetime structure before the bounce}\label{before-bounce}

We showed in Figure \ref{fig:4} the existence of a trapped region, that is $\theta_{\mathrm{in}} \theta_{\mathrm{out}} > 0$ where $\theta _{\mathrm{in}} < 0$ and $\theta _{\mathrm{out}} < 0$, for $t < 0$. This region is bounded from below by  inner $(X_{-})$ and outer $(X_{+})$ trapping horizons for which $\theta_{\mathrm{out}} =0$. As we approach the bounce, both horizons get closer and at
\begin{equation}\label{tast}
 t = t_{\ast} = - \left( \sqrt{3}\frac{G m_0}{c^3 t_0} + \sqrt{\left(\sqrt{3}\frac{G m_0}{c^3 t_0}\right)^2 - 1}\right),
 \end{equation}
 only a trapping horizon exist for $X_{-} = X_{+} = 3$. If we choose $m_{0} = 50 \; M_{\odot}$, and $t_{0} = 5 \times 10^{-5}$ s, then $t_{\ast} = - 16.9$.
 
\begin{figure}[t]
\includegraphics[width=8cm]{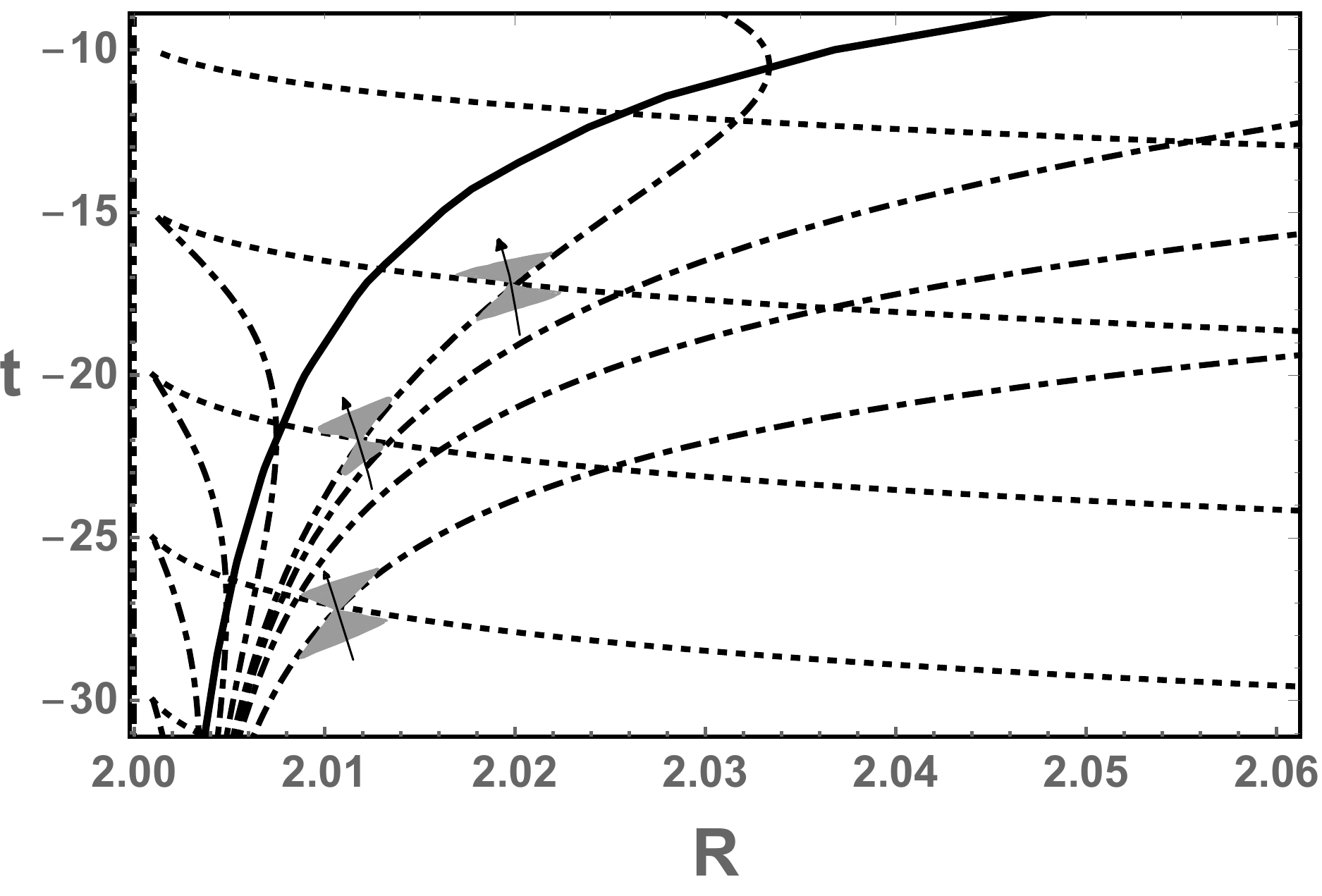}
\caption{\label{fig:10}Light cone structure in McVittie spacetime for a bouncing  cosmological model in the region $t<0$. The dotted curves represent the null ingoing geodesics while the dot dashed curves the null outgoing geodesics. The regions shaded in grey show some light cones and the black arrow indicates the future direction.}
\end{figure}

 In Section \ref{th-nullgeo}, we proved that $X_{+}$ is a null infinity ($X_{+} \rightarrow \infty$ for $t \rightarrow -\infty$). The inner horizon $X_{-}$, defined for $2 < X_{-} \le 3 $, covers the singularity and encloses a trapped region. We illustrate this situation in Figure \ref{fig:10} where the light cone structure for small and negative values of $t$ is plotted. The light cones have the trapping horizon in their local future . Both ingoing and outgoing null rays that cross $X_{-}$ enter the trapped region and are unable to turn around and escape. This horizon, thus, acts as a one way membrane, hiding the singularity at $x = 2 $. This analysis leads us to conclude that in the time interval $- \infty < t < t_{\ast} $ the solution contains a black hole.



\section{Discussion}\label{sec:discussion}

Given the analysis of the causal structure of the spacetime offered in the previous sections, we now provide a description of the evolution of the solution through cosmic time.

A black hole is present since the beginning of the contracting phase. The inner trapping horizon $X_{-}$ increases its radius as the contraction gathers pace. The range of values of the radial coordinate for  $X_{-}$ is $2 < X_{-} \le 3 $. Ingoing and outgoing null geodesics that cross the surface $X_{-}$, enter the trapped zone of the spacetime interior to the black hole. Close to the bounce, at $t_{\ast}$ (see Eq. \eqref{tast}), the inner $X_{-}$ and outer $X_{+}$ trapping horizons merge and the black hole ceases to exist. This situation lasts for a short time. Trapping horizons appear again right before the bounce, and vanish right after it. These horizons are absent in other McVittie models (for instance, compare Figures \ref{fig:1} and \ref{fig:1-1} with \ref{fig:4}). We associate these surfaces with the peculiarities of the cosmological background model, and more specifically to the presence of the bounce. There is no salient feature associated with these horizons: there, null geodesics just change their convergence properties.

\begin{figure}[t]
\includegraphics[width=8cm]{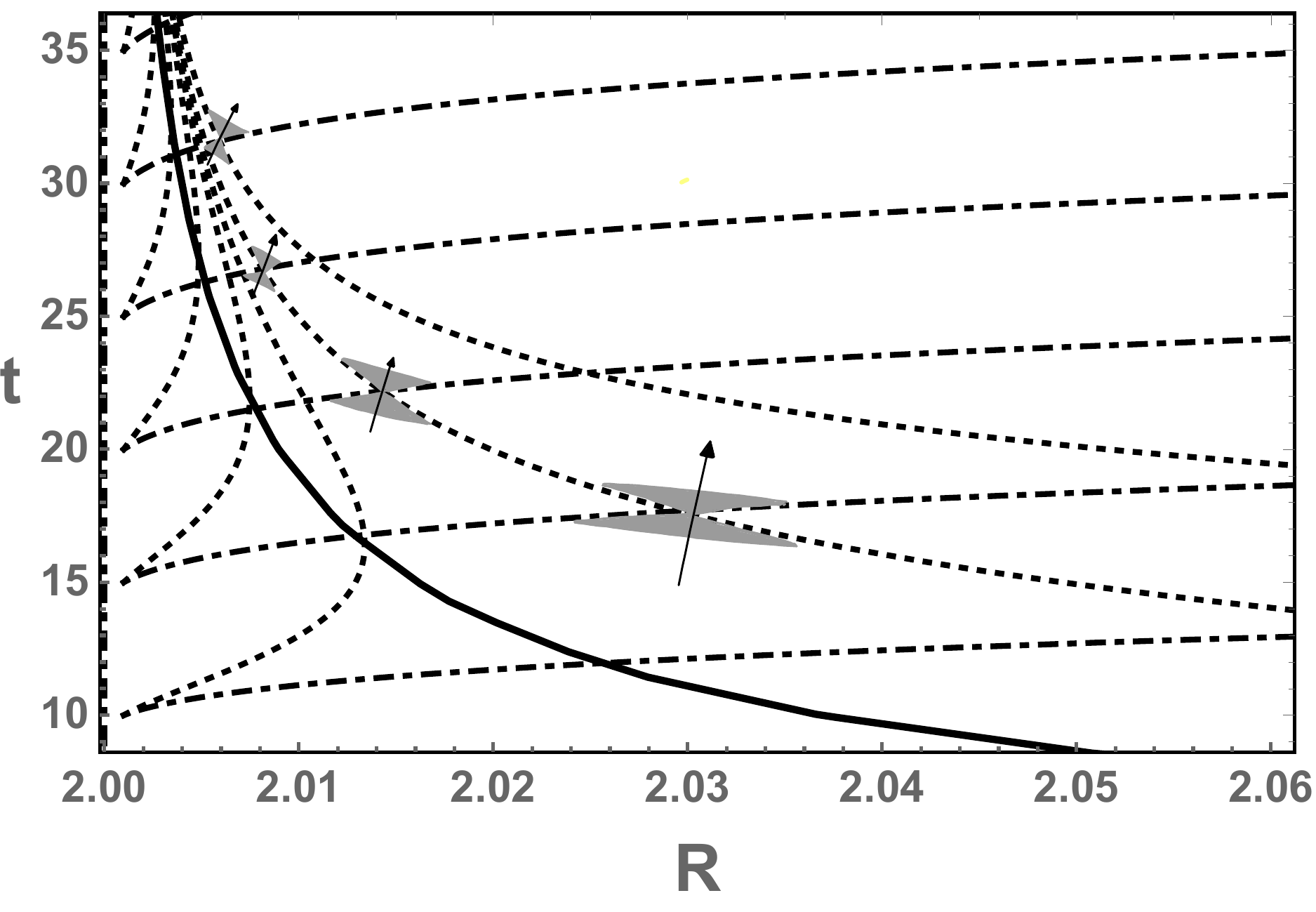}
\caption{\label{fig:13lc} Light cone structure in McVittie spacetime for a bouncing  cosmological model in the region $t>0$. The dotted curves represent the null ingoing geodesics while the dot dashed curves the null outgoing geodesics. The grey shadow regions show some light cones and the black arrow indicates the future direction.}
\end{figure}

Afterwards, the universe begins to expand and an inner $X_{-}$ and outer $X_{+}$ trapping horizons appear. For $t \rightarrow \infty$, the outer horizon becomes a FLRW null infinity, while $X_{-}$ becomes an event horizon. Hence, as the universe expands a black hole starts to form. We show in Figure \ref{fig:13lc} the light cone structure for the spacetime after the bounce. We see that the surface $x= 2$ is not in the local future of those light cones: in the process of black hole formation (the event horizon is not settled down) some geodesics are able to escape from the central source. As the universe expands, however, those outgoing geodesics that start at $x < X_{-}$ have a smaller and smaller slope. In the limit $t \rightarrow \infty$, ${dx/dt} \rightarrow  0$ for outgoing geodesics, as  computed from Eq. \eqref{out-in}. This implies that outgoing null geodesics cannot leave the surface $x = 2 $ and the region contained by such boundary becomes trapped. In the distant future, the McVittie solution for a bouncing cosmological model harbors a black hole.


\begin{figure}[t]
\includegraphics[width=9.5cm]{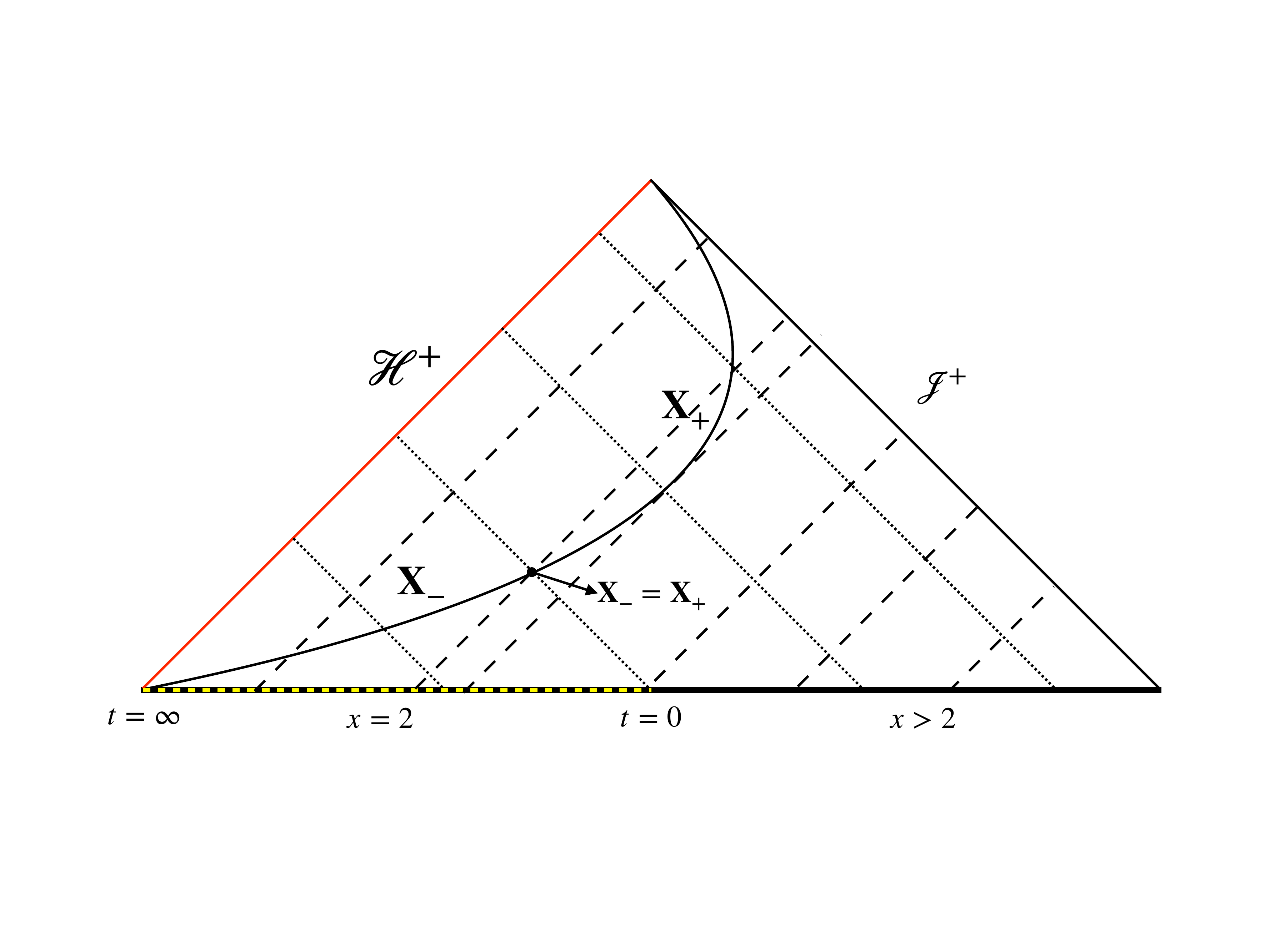}
\caption{\label{fig:13}Qualitative Penrose diagram of the McVittie spacetime for a bouncing cosmological model in the expanding region ($t >0$). The dotted lines represent null ingoing geodesics while the dashed lines null outgoing geodesics. Here, $\mathcal{H^{+}} = X_{-} \rightarrow 2, \; \; t \rightarrow \infty$, and $\mathcal{J}^{+}$ is the future null infinity.}
\end{figure}

\begin{figure}[t]
\includegraphics[width=9.5cm]{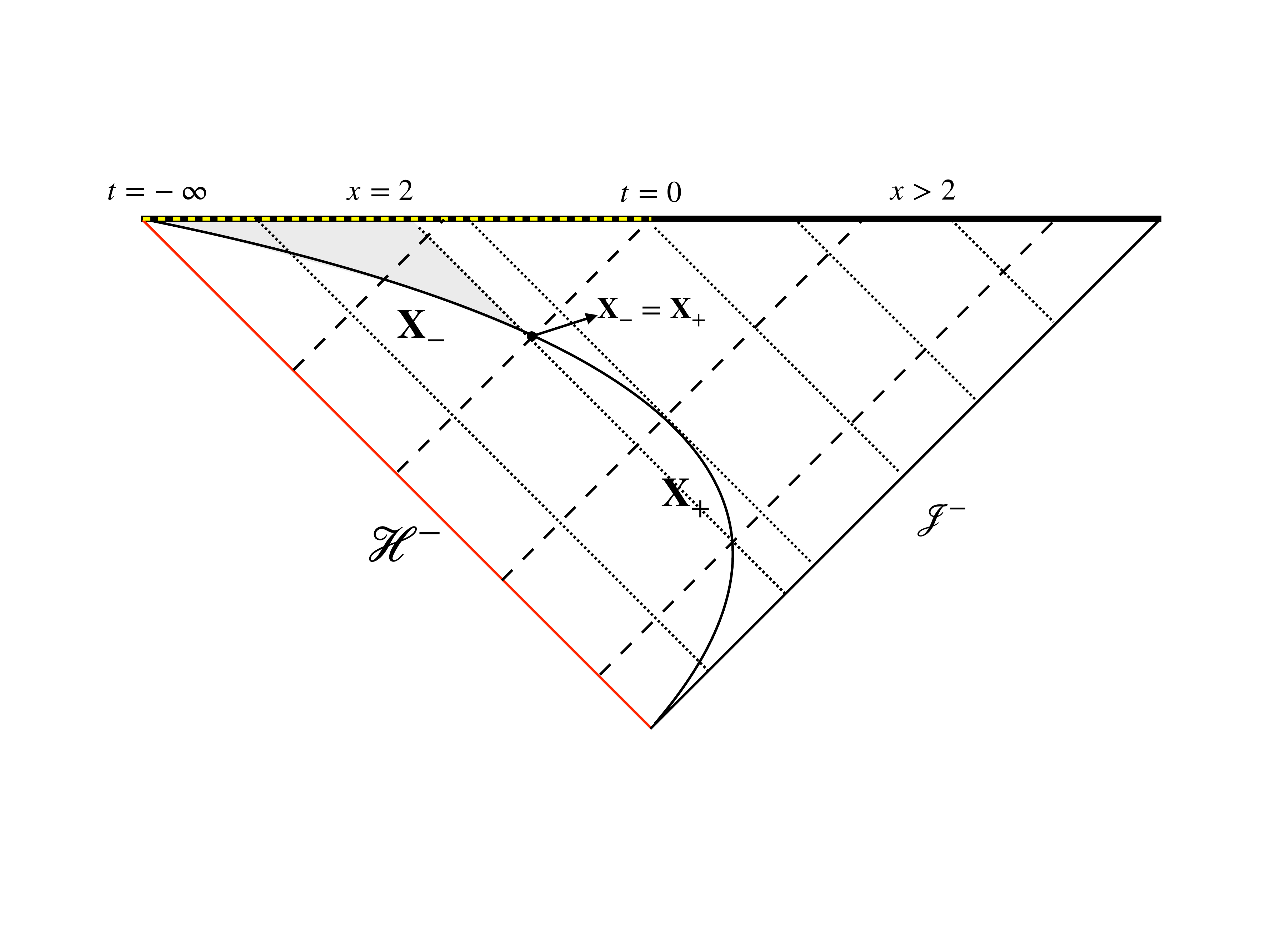}
\caption{\label{fig:14}Qualitative Penrose diagram of the McVittie spacetime for a bouncing cosmological model in the contracting region ($t < 0$). The dotted lines represent null ingoing geodesics while the dashed lines null outgoing geodesics. Here, $\mathcal{H^{-}} = X_{-} \rightarrow 2, \; \; t \rightarrow- \infty$, and $\mathcal{J}^{-}$ is the past null infinity.}
\end{figure}

We show in Figures \ref{fig:13} and \ref{fig:14} qualitative Penrose diagrams of the McVittie spacetime for a bouncing cosmological model in the expanding and contracting phase, respectively. The dotted lines represent the trajectories of ingoing null geodesics and the dashed lines the trajectories of outgoing null geodesics. The black thick curve displays the trapping inner ($X_{-}$) and outer ($X_{+}$) horizons. The  straight red line named $\mathcal{H^{+}}$ ($\mathcal{H^{-}}$) represents  $X_{-} \rightarrow 2, \; \; t \rightarrow \infty$ ($X_{-} \rightarrow 2, \; \; t \rightarrow - \infty$). There is a horizontal straight black line that characterises the region where the bounce occurs, that is  $t = 0$ and $x > 2$. To the left, a dashed black and yellow line represents the singular surface $x = 2$, $t$ finite. As usual, $\mathcal{J^{+}}$ ($\mathcal{J^{-}}$) denotes the future (past) null infinity.

In the expanding phase, all ingoing geodesics, regardless of the initial conditions, reach the surface $\mathcal{H^{+}}$. Those outgoing null geodesics that start at the bounce extend to $\mathcal{J^{+}}$. There are also future directed outgoing radial null geodesics that emerge from the singular surface $x = 2$, $t$ finite (this includes $(t,x) = (0,2)$). The latter evidences the existence of a  spacelike naked singularity that lasts up to the formation of the black hole which corresponds  to $\mathcal{H^{+}}$ in Figure \ref{fig:13}.

As shown in Section IV A, in the contracting phase, ingoing and outgoing null geodesics reverse their character. Some ingoing geodesics end up in the singular surface while some others make it to the bounce. The same fate share outgoing null geodesics; all of them begin in $\mathcal{H^{-}}$. A black hole is present since the beginning of the contracting phase until $X_{-} = X_{+}. $The zone shaded in grey indicates the black hole region.

\section{Conclusions}

    In this work the causal structure of McVittie spacetime for a bouncing cosmological background is analyzed. The location of the trapping horizons is computed, and the trajectories of null radial ingoing and outgoing null geodesics through cosmic time are obtained by numerical integration. A detailed study of the asymptotic behaviour of the metric is provided. Our main result is that the solution represents a dynamical black hole since the beginning of the contracting phase up to shortly before the bounce, and also in the distant future. Just before the bounce, the inner and outer trapping horizons merge and the black hole character of the solution is lost. After the bounce, the central inhomogeneity starts to act again, and for large and positive values of the cosmic time, a black hole is formed. Thus, we see that the global dynamical state of the universe directly affects the conditions for the existence of black holes. In particular, for a contracting universe, black hole solutions are possible up to certain minimum scales.

Unlike all other McVittie models analyzed in the literature, there is no cosmological big singularity in the present metric. In fact, the solution admits trajectories that never encounter a singularity, that is, they are geodesically complete. This peculiar feature of the model is related to the occurrence of the bounce.

This work is a first step towards a better understanding of black holes embedded in a bouncing cosmological background; the current solution does not take into account the accretion of cosmological fluid by the central source. The Generalized McVittie metric naturally incorporates this effect by assuming an energy-momentum tensor of an imperfect fluid \cite{far15}. It remains an open issue whether the Generalized McVittie spacetime might contain a black hole for a bouncing cosmological background. We shall explore this issue in a future work.


\begin{acknowledgments}
This work was supported by the Argentine agency CONICET (PIP 2014-00338) and the Spanish Ministerio de Ciencia e Innovación (MICINN) under grant PID2019-105510GBC31 and through the ”Center of Excellence María de Maeztu 2020-2023” award to the ICCUB (CEX2019-000918-M). G.E.R. acknowledges support from the \emph{Coordena\c c\~ao
de Aperfei\c coamento de Pessoal de N\'ivel Superior-Brasil
(CAPES)-Codigo de Financiamento 001}. Both D.P. and G.E.R. are very grateful to the Department of Physics of UERJ for kind hospitality. We are very grateful to two anonymous referees for their insightful comments.
\end{acknowledgments}




\bibliography{apssamp}

\end{document}